\documentclass[twocolumn,superscriptaddress,amsmath,amssymb]{revtex4-2}

\usepackage{color}
\usepackage{graphicx}
\usepackage{subfigure}
\usepackage{booktabs}
\usepackage{dcolumn}
\usepackage{bm}
\usepackage{amsmath}
\usepackage{verbatim}

\newcommand{\beq}{\begin{eqnarray}}
\newcommand{\eeq}{\end{eqnarray}}

\begin{document}

\title{Shear hardening in frictionless amorphous solids near the jamming transition}
\author{Deng Pan}
\affiliation{CAS Key Laboratory of Theoretical Physics, Institute of Theoretical Physics, Chinese Academy of Sciences, Beijing 100190, China}

\author{Fanlong Meng}
\author{Yuliang Jin}
\email{yuliangjin@mail.itp.ac.cn}
\affiliation{CAS Key Laboratory of Theoretical Physics, Institute of Theoretical Physics, Chinese Academy of Sciences, Beijing 100190, China}
\affiliation{School of Physical Sciences, University of Chinese Academy of Sciences, Beijing 100049, China}
\affiliation{Wenzhou Institute, University of Chinese Academy of Sciences, Wenzhou, Zhejiang 325000, China}

\date{\today}

\begin{abstract}

The jamming transition, generally manifested by a rapid increase of rigidity under compression (i.e., compression hardening), is ubiquitous in amorphous materials.   Here we study shear hardening in deeply annealed frictionless packings generated by numerical simulations, reporting critical scalings absent in compression hardening. We demonstrate that hardening is a natural consequence of shear-induced memory destruction. Based on an elasticity theory, we reveal two independent microscopic origins of shear hardening: (i) the increase of the interaction bond number and (ii) the emergence of anisotropy and long-range correlations in the orientations of bonds – the latter highlights the essential difference between compression and shear hardening. Through the establishment of physical laws specific to anisotropy, our work completes the criticality and universality of jamming transition, and the elasticity theory of amorphous solids.

\end{abstract}

\maketitle

\newpage

\section*{Introduction}

The jamming transition occurs in a wide range of soft materials, ranging from granular matter to colloidal suspensions to glasses. As a non-equilibrium, athermal phase transition, its criticality is specified by a set of scaling laws~\cite{makse2000packing, OHern2003, Charbonneau2014a, Goodrich2016}. In particular, the scaling relationship (under the harmonic approximation of the inter-particle interaction)  $G \sim \Delta Z$, between the shear modulus $G$ of the jammed phase and the excess coordination number $\Delta Z$, has been derived by the mean-field replica glass theory in infinite dimensions~\cite{Yoshino2014} and microscopic elasticity theories~\cite{Wyart2005,  Zaccone2011a}, with a promising agreement to data from isotropic compression simulations~\cite{OHern2003}. Here we demonstrate that this relationship cannot fully account for the shear hardening behavior -- the influence of fabric anisotropy (quantified by the macroscopic friction  coefficient $\mu = \sigma/P$, which is the ratio of shear stress $\sigma$ to pressure $P$) is significant.
We report numerically a new scaling law unique to shear, $\mu \sim \sigma^{\beta}$ with $\beta \approx 0.25$,  between $\mu$ and 
$\sigma$, and derive theoretically an additional anisotropic contribution to the elasticity, $G_{\rm AI} \sim \mu^2$, verified by our simulation data.

{\it Shear hardening}, meaning that $G$ increases by shear, is closely related to several other interesting phenomena. The vestige of shear hardening in the unjammed phase is the phenomenon of {\it shear jamming} (the onset of rigidity under a fixed volume shear deformation) observed in many experiments~\cite{Bi2011, Ren2013, Sarkar2013, wang2018microscopic, zhao2022ultrastable} and simulations~\cite{Kumar2016, Bertrand2016, Vinutha2016, BaityJesi2017, Jin2018, seto2019shear, xiong2019comparison, jin2021jamming, Babu2021, Kawasaki2020, otsuki2020shear, Singh2020, giusteri2021shear, babu2022criticality}. 
Shear hardening is generally accompanied by increasing  pressure under fixed-volume conditions, which implies that under fixed-pressure conditions a {\it dilatancy effect} would occur~\cite{Reynolds1885, Rowe1962, Rao2008}.
Recent studies have shown the possibility of complete decoupling between friction and shear jamming/dilatancy~\cite{Urbani2017, Jin2018, jin2021jamming, Babu2021}.
In this study we show that friction is also not essential for shear hardening.
Due to the lack of internal friction, the conventional sawtooth model~\cite{bolton1986strength} cannot be applied anymore, and frictionless mechanisms are demanded. To this end, here we provide a phenomenological explanation based on the generalized jamming phase diagram~\cite{Babu2021}, and reveal microscopic origins of frictionless shear hardening from the elasticity theory of amorphous solids~\cite{Maloney2004,  lemaitre2006sum, Karmakar2010athermal, Zaccone2011, Zaccone2011a, Wyart2005}.
 Our simulation and theoretical results demonstrate that fabric anisotropy makes important contributions to shear hardening, which  resembles a similar connection between anisotropy and shear jamming~\cite{Vinutha2016}. 
Note that the discussed shear hardening occurs in the solid phase (above jamming), which shall not be confused with {\it shear thickening} that presents in non-Newtonian fluids (below jamming) such as dense suspensions~\cite{ness2022physics}.


Shear hardening has been directly observed in previous studies~\cite{Boschan2016, Kawasaki2020, zhao2022ultrastable, pan2022nonlinear}. Simulations report a hardening regime on the stress-strain curve, following elastic and softening regimes~\cite{Boschan2016, Kawasaki2020}.
These previous studies~\cite{Boschan2016, Kawasaki2020} consider mechanically trained packings,  which  have effectively  a moderate degree of annealing. The shear hardening scaling regime in such systems is limited (about one decade) due to the interruption of yielding (note that the yield stress $\sigma_{\rm Y}$ vanishes in the rapid quench limit~\cite{Heussinger2009}), making a reliable estimation of the  scaling exponents  difficult.
To overcome this issue, here we simulate ultra-stable packings annealed by an efficient swap algorithm~\cite{kranendonk1991computer, grigera2001fast, gutierrez2015static, Berthier2016, Ninarello2017}. 
The yield stress of our ultra-stable packings is significantly larger than that of mechanically trained packings~\cite{pan2022nonlinear}, and consequently the shear hardening scaling regime is  extended up to about eight decades. This advantage allows us to accurately determine shear hardening exponents, which turn out to be independent of the degree of annealing.


\section*{Results}

\begin{figure}[!htb]
  \centering
  \includegraphics[width=0.9\linewidth]{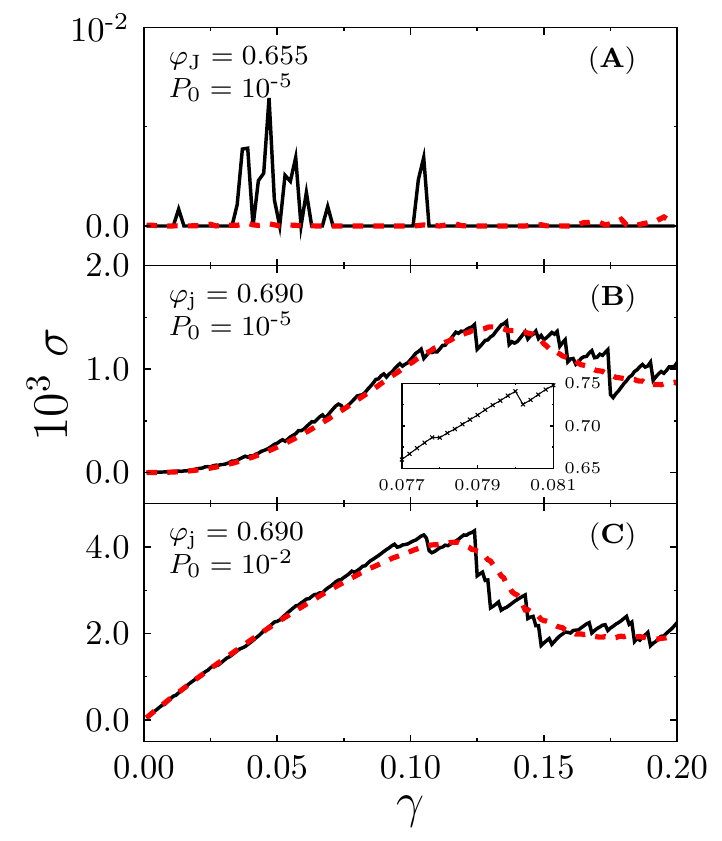}
  \caption{{Stress-strain curves.} Plotted are stress-strain curves of  {(A) a rapidly quenched 
  system near the jamming transition, (B) a deeply annealed system near the jamming transition, and (c) a deeply annealed system over-compressed well above the jamming transition.
  } 
  Data are obtained from constant volume AQS simulations of 3D frictionless soft spheres (SSs).
  The {dashed} and {solid} lines represent averaged and single-realization curves respectively. {The inset in panel (B) is an  enlarged view of the single-realization curve.}}
  \label{fig:singleAveStressCurve}
\end{figure}

\begin{figure}[!htb]
  \centering
  \includegraphics[width=0.9\linewidth]{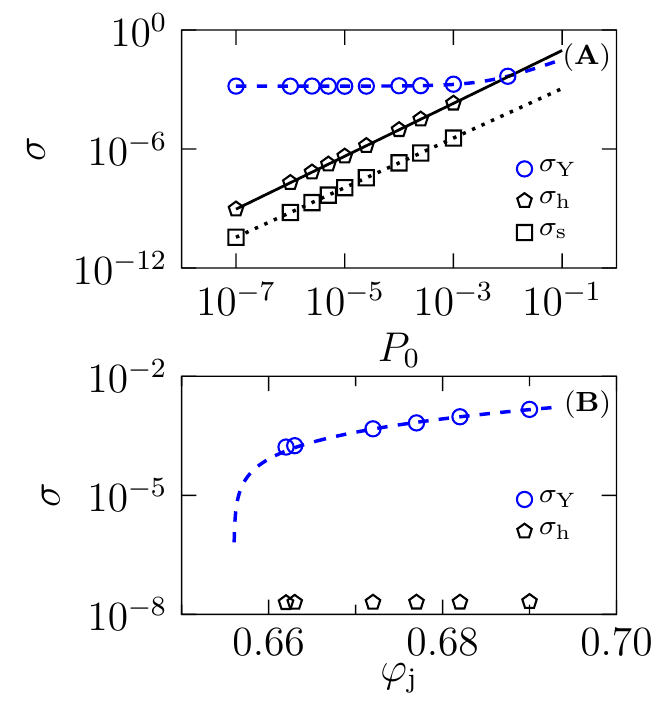}
  \caption{
  {
  Dependence of several characteristic stresses on $P_0$ and $\varphi_{\rm j}$.
  (A) The yielding stress $\sigma_{\rm Y}$, the onset stress $\sigma_{\rm h}$ of shear hardening, and the onset stress $\sigma_{\rm s}$ of shear softening are plotted as functions of the unstrained pressure $P_0$, with a fixed $\varphi_{\rm j} = 0.69$. The dashed line is the fitting curve $\sigma_{\rm Y}(P_0) = 1.48\times10^{-3} + 0.317 P_0$. The solid and dotted lines are the relations $\sigma_{\rm h}(P_0) = 2.0 \times P_0^{4/3}$ and $\sigma_{\rm s}(P_0) = 0.02 \times P_0^{5/4}$ (see Fig.~\ref{fig:collapse}). 
  The two lines $\sigma_{\rm Y}(P_0)$ and $\sigma_{\rm h}(P_0)$ intersect at {$P_0^* = 1.13\times10^{-2}$}, above which shear hardening is non-observable. 
  (B) The $\varphi_{\rm j}$-dependence of $\sigma_{\rm Y}$  and $\sigma_{\rm h}$, with a fixed $P_0 = 10^{-6}$. The dashed line is the fitting curve $\sigma_{\rm Y}(\varphi_{\rm j}) = 0.0175 (\varphi_{\rm j} - \varphi_{\rm J}) + 0.736 (\varphi_{\rm j} - \varphi_{\rm J})^2$, where
  $\varphi_{\rm J} \approx 0.655$. 
  It is known that   $\sigma_{\rm Y}$  vanishes linearly with $\varphi_{\rm j} - \varphi_{\rm J}$ in the vicinity of $\varphi_{\rm J}$~\cite{Babu2021}; here a quadratic correction is added to account for the  deviation from this linear critical scaling in the 
  large $\varphi_{\rm j}$ regime. 
 }
  }
  \label{fig:sigYsigH}
\end{figure}

\begin{figure*}[!htb]
  \centering
  \includegraphics[width = 0.85 \linewidth]{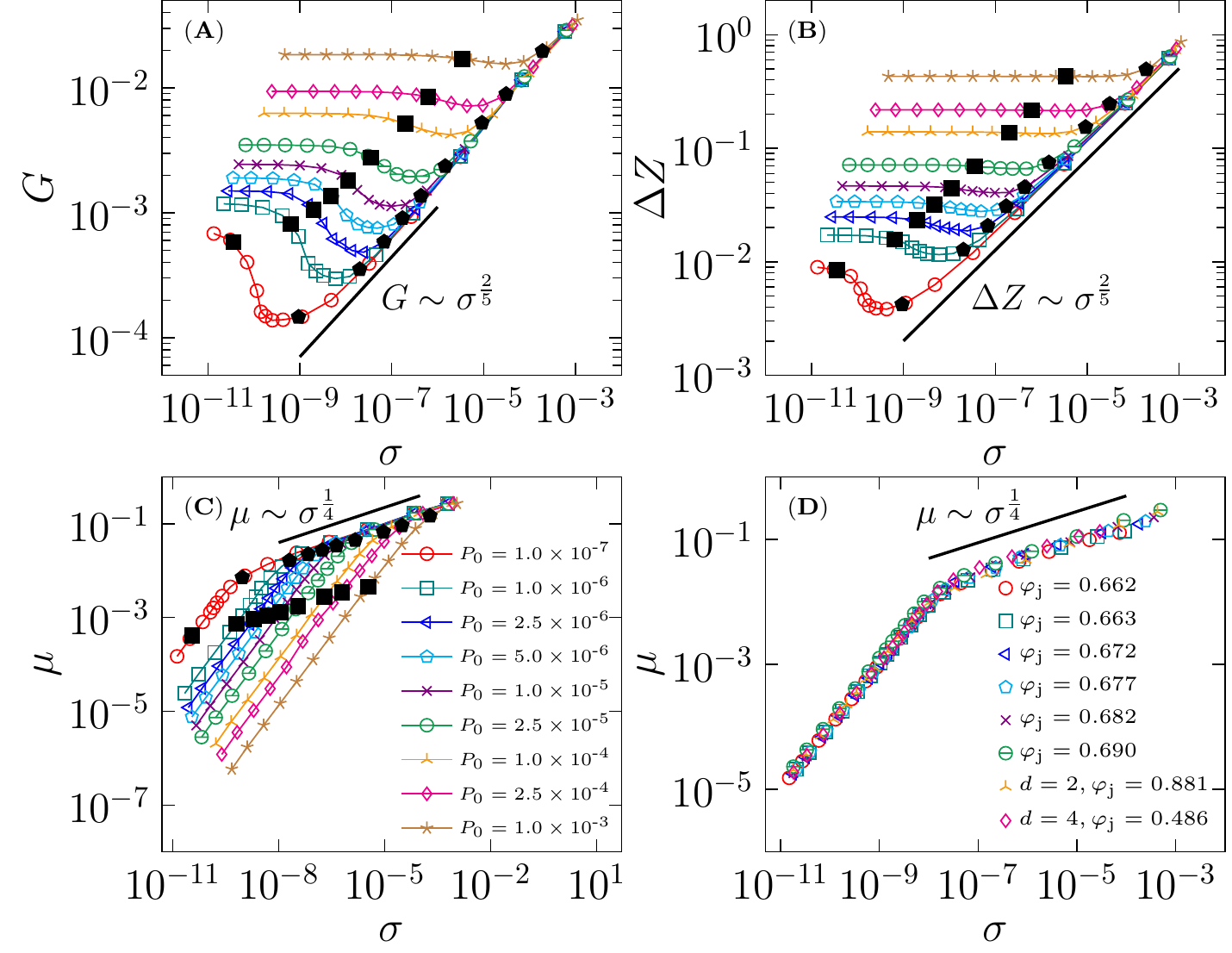}
  \caption{{Shear hardening scalings.}
  Simulation results in 3D of (A) the shear modulus $G$,
  (B) the excess coordination number $\Delta Z$ and (C) the macroscopic friction coefficient $\mu$ as functions of stress $\sigma$, for a few different $P_0$. The solid lines represent scaling laws in the shear hardening regime, $G\sim \sigma^{2/5}$, $\Delta Z \sim \sigma^{2/5}$ and $\mu \sim \sigma^{1/4}$. The crossovers at $\sigma_{\rm s}$ and $\sigma_{\rm h}$, which are determined in Fig.~\ref{fig:collapse}, are marked by black solid squares and pentagons respectively.
  (D) Universal relationship between $\mu$ and $\sigma$
  in 3D samples  prepared by cyclic shear ($\varphi_{\rm j} = 0.662$), cyclic compression ($\varphi_{\rm j} = 0.663$), and swap thermal annealing ($\varphi_{\rm j} = 0.672, 0.677, 0.682, 0.690$), as well as in two and four dimensions ($P_0 = 10^{-6}$ in all cases).
  }
  \label{fig:results_stress}
\end{figure*}

\subsection*{Conditions for shear hardening} As demonstrated by the linear plots in  Fig.~\ref{fig:singleAveStressCurve}, the stress-strain curves, averaged over samples,  typically display three kinds of behavior under constant-volume athermal quasi-static shearing (AQS).
(i) Rapidly quenched systems with the minimum jamming density (see 
Materials and Methods),  $\varphi_{\rm j} \approx \varphi_{\rm J}$, behave like {\it plastic flows} in the  zero pressure limit $P_0 \to 0$,  where $P_0 \equiv P(\gamma = 0)$ is the pressure of unstrained configurations: $\sigma(\gamma) \approx 0$ and  $G(\gamma) \approx 0$  (see Fig.~\ref{fig:singleAveStressCurve}A). On the stress-strain curve of an individual sample obtained in a single realization of simulation, the coexistence of solid-like ($\sigma>0$ and $G>0$) and liquid-like ($\sigma=G=0$) states can be identified~\cite{Heussinger2009}.
(ii)  Shear hardening, manifested in the increasing function of $G(\gamma)$, appears in deeply annealed systems with $\varphi_{\rm j} \gg \varphi_{\rm J}$, at small $P_0$ (see Fig.~\ref{fig:singleAveStressCurve}B). {Note that in this case, there is a narrow {\it shear softening}  regime (i.e., $G(\gamma)$ is a descending function) before shear hardening, which can only be identified in log scales (see Fig.~\ref{fig:results_stress} and Supporting Information SI Fig.~S1).} 
(iii) If the systems are over-compressed { well above the jamming transition,
then shear hardening disappears and only shear softening}
is observed  independent of $\varphi_{\rm j}$ (see Fig.~\ref{fig:singleAveStressCurve}C). 

From the above analysis, one finds that shear hardening emerges near the jamming transition ($P_0 \to 0$) in 
deeply annealed packings ($\varphi_{\rm j} \gg \varphi_{\rm J}$),
under constant-volume AQS. 
These two conditions are demonstrated more quantitatively in Fig.~\ref{fig:sigYsigH}. 
The shear hardening behavior can be only observed in a stress window $\sigma_{\rm h} < \sigma < \sigma_{\rm Y}$, where $\sigma_{\rm h}$ is the   onset stress of hardening (see below for how $\sigma_{\rm h}$ is determined).
For a fixed  $\varphi_{\rm j}$ with increasing  $P_0$,  $\sigma_{\rm h}$ increases faster than $\sigma_{\rm Y}$, as shown in  Fig.~\ref{fig:sigYsigH}A. 
Above $P_0^*$, where $\sigma_{\rm h} (P_0^*)\approx \sigma_{\rm Y} (P_0^*)$,  the shear hardening regime disappears (The finite-size analysis is in SI, see Fig.~S5).  
For a fixed $P_0$,  $\sigma_{\rm h}$ is nearly independent of $\varphi_{\rm j}$ (see Fig.~\ref{fig:sigYsigH}B), while $\sigma_{\rm Y}$ decreases with decreasing $\varphi_{\rm j}$ and vanishes when $\varphi_{\rm j} \to \varphi_{\rm J}$~\cite{Liu1998, Heussinger2009} 
(a more detailed study on $\sigma_{\rm Y}(\varphi_{\rm j})$ can be found in Ref.~\cite{Babu2021}). Thus the scaling regime of shear hardening also vanishes as  $\varphi_{\rm j} \to \varphi_{\rm J}$.
The finite-size effects are discussed in SI Sec. 5.
In this study, we do not consider other factors, such as  finite temperatures or finite shear rates~\cite{das2019scaling}.

\begin{figure}[!htb]
  \centering
  \includegraphics[width=0.95\linewidth]{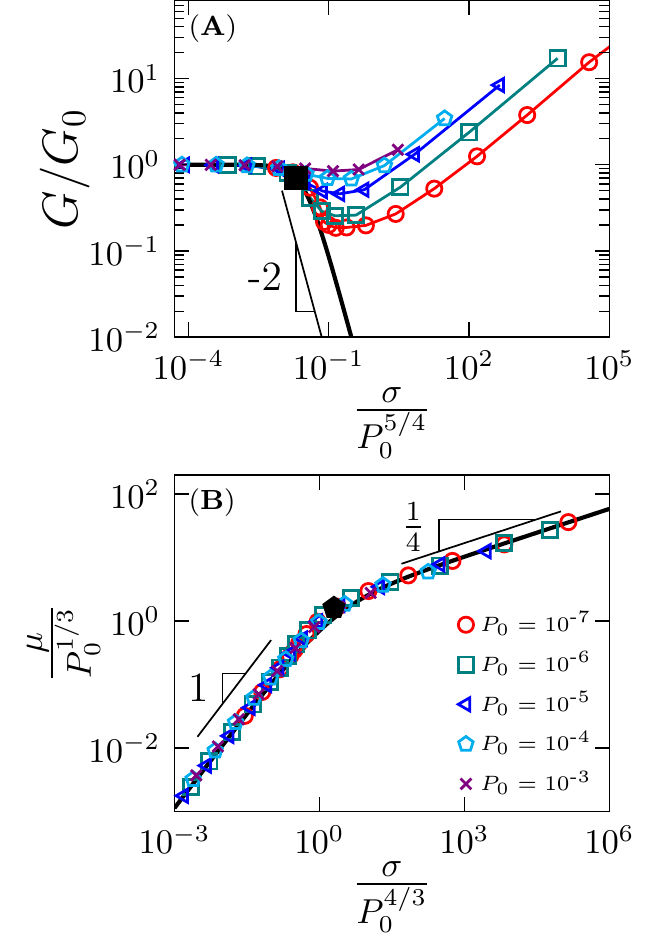}
  \caption{{Data collapsing in shear softening and hardening regimes.}
  Rescaled plots of (A) shear modulus $G$ and (B) macroscopic friction $\mu$ as functions of stress $\sigma$, where $P_0$ and $G_0$ are unstrained pressure and shear modulus.
  The crossover stresses are numerically defined as $\sigma_{\rm s}/P_0^{5/4} = 2\times10^{-2}$ (solid black square) in (A) and $\sigma_{\rm h}/P_0^{4/3} = 2$ (solid black pentagon) in (B).
  The bold lines represent the master curves $y = \frac{1}{1 + (32.7 x)^2}$ and $y = \frac{1.13 x}{1+(0.52  x)^{0.75}}$ in (A) and (B) respectively .
  }
  \label{fig:collapse}
\end{figure}

\subsection*{Critical scaling laws} Below we focus on deeply annealed samples with a large jamming density $\varphi_{\rm j}=0.69$, which are quenched from dense equilibrium liquids at $\varphi_{\rm eq} = 0.643$.
Based on the log-log plots in Fig.~\ref{fig:results_stress}, we identify consecutively (before yielding) the elastic, shear softening and shear hardening regimes, separated by two crossover points at $\sigma_{\rm  s}$ and $\sigma_{\rm  h}$. 

The data of shear modulus in the elastic and shear softening regimes can be described by a scaling function $G(P_0, \sigma)/G_0 = \mathcal{G}(\sigma/\sigma_{\rm s})$, where $G_0 = G(P_0, 0)$ is the unstrained shear modulus, $\mathcal{G}(x \to 0) = 1$, $\mathcal{G}(x \to \infty) = x^{-2}$  and  $\sigma_{s} \sim P_0^{5/4}$ (see Fig.~\ref{fig:collapse}A).
{The scaling function $\mathcal{G}(x)$ was initially proposed in Ref.~\cite{Nakayama2016}
to describe both linear and softening behavior in rapidly quenched packings. 
We find that $\mathcal{G}(x)$ agrees well with our  data of deeply annealed systems in the linear and softening regimes, but 
deviate from the data
starting from $\sigma = \sigma_{\rm h}$ which is the onset point of shear hardening. The scaling function  $\mathcal{G}(x)$  allows us to determine numerically the crossover stress $\sigma_{\rm s} = c P_0^{5/4}$, where the prefactor $c$ is chosen to be 0.02. The $\sigma_{\rm s}$ estimated in this way (squares) reasonably separates linear and softening behavior, as can be seen in Fig.~\ref{fig:results_stress}A. From $\mathcal{G}(x)$, the scaling behavior of linear and softening regimes can be extracted. In the linear regime ($\sigma \ll \sigma_{\rm s}$ or $x = \sigma /\sigma_{\rm s} \to 0$),  $\mathcal{G}(x \to 0) = 1$, which means that the shear modulus is simply a constant $G = G_0$. In the softening regime ($\sigma \gg \sigma_{\rm s}$ or $x  \to \infty$), $G(x \to \infty) = x^{-2}$ which means that $G \sim P_0 \gamma^{-2/3}$ (where we have used the well know scaling $G_0 \sim P_0 ^{1/2}$ for un-strained systems~\cite{OHern2003}).  These results are fully consistent with the findings in Ref.~\cite{Nakayama2016}.
Our data 
show that the presence of a strong shear hardening effect does not change the scalings in linear and softening regimes. 
It is of interest to provide a theoretical explanation of 
 the softening exponent $-2$ in  $G(x \to \infty) = x^{-2}$  in future studies.}


The deviation of the simulation data from the master curve $\mathcal{G}(x)$ in Fig.~\ref{fig:collapse} is followed by the shear hardening behavior. The scaling of   $\sigma_{\rm h}$ is determined  based on the rescaled plot of macroscopic friction $\mu$ in Fig.~\ref{fig:collapse}B, with a scaling function $\mu(P_0, \sigma)/\mu_{\rm  h} = \mathcal{U}(\sigma/\sigma_{\rm h})$, where $\mathcal{U}(x \to 0) \sim x$, $\mathcal{U}(x \to \infty) \sim x^{1/4}$, $\sigma_{\rm h} \sim P_0^{4/3}$ and $\mu_{\rm h}\sim P_0^{1/3}$.
Figure~\ref{fig:collapse}B shows that elastic and softening data follow a universal scaling $\mu \sim \sigma$.
Previously, Ref.~\cite{Kawasaki2020} reports two scalings $\mu \sim \gamma$ and $\mu \sim \gamma^{1/2}$ respectively in the elastic and softening regimes for a two-dimensional model, which, together with the shear softening scaling $\sigma \sim \gamma^{1/2}$, give a consistent result $\mu \sim \sigma$.

In the shear hardening regime, $\sigma_{\rm h} < \sigma < \sigma_{\rm Y}$, the system's behavior is characterized by two scaling laws. {Numerical fitting gives $\mu \sim \sigma^{\beta}$ with $\beta = 0.248 \pm 0.006$ and $\Delta Z \sim \sigma^{\nu}$ with $\nu = 0.411 \pm 0.005$ in the shear hardening regime (Fig.~\ref{fig:results_stress}).} The scaling of $\Delta Z$ is consistent with the ansatz $\Delta Z \sim \sigma^{2/5}$ proposed  in Ref.~\cite{Goodrich2016}.
Note that this scaling is examined in~\cite{Goodrich2016} by looking at the stress variance  $\sigma^2$ of isotropically compressed configurations with $\varphi_{\rm j} = \varphi_{\rm J}$, where the mean stress is zero. Here we provide direct verification of the ansatz, thanks to the emergence of shear hardening in deeply annealed packings ($\varphi_{\rm j} \gg \varphi_{\rm J}$). The scaling $\mu \sim \sigma^{1/4}$ (or equivalently $P \sim \sigma^{3/4}$), is absent in the framework of Ref.~\cite{Goodrich2016}, and to our knowledge, has never been reported previously.
Based on the elasticity theory (see below for details), we derive that $G$ is a linear combination of $G_{\rm I} \sim \Delta Z$ and $G_{\rm AI} \sim \mu^2$ ,  which, together with the above two scalings, gives $G \sim \sigma^{2/5}$ in the jamming limit { where $\sigma \to 0$} (see Fig.~\ref{fig:results_stress}A for a comparison to the numerical data). While it is straightforward to translate these scalings, $G \sim \Delta Z \sim \sigma^{2/5}$ and $\mu \sim \sigma^{1/4}$, from the stress-dependent forms into strain-dependent forms based on the relation $G = d\sigma/d\gamma$, one should be cautious about the influence of plasticity in the measurement of stress and modulus~\cite{Nakayama2016}.

It can be shown that the shear hardening scaling exponents are independent of the preparation method (thermal annealing or mechanical training),  the degree of annealing represented by $\varphi_{\rm j}$,
and the dimensionality in $d=2, 3, 4$  dimensions (see SI Fig.~S2, S3 and S4). In particular, the data of $\mu$ versus $\sigma$ collapse remarkably without any shifting (see Fig.~\ref{fig:results_stress}D).

\begin{figure*}[!htb]
  \centering
  \includegraphics[width=0.8\linewidth]{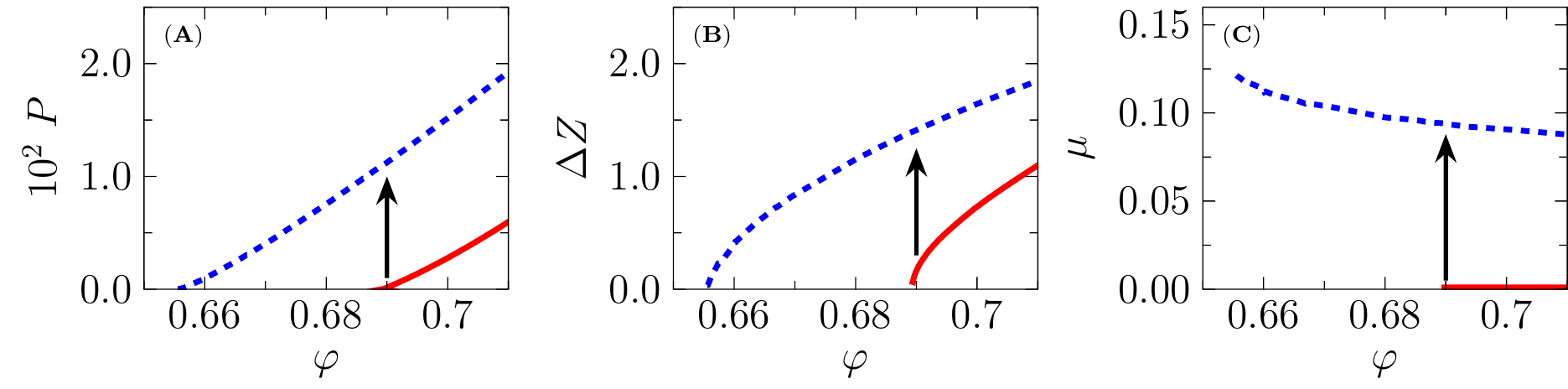}
  \caption{{Equations of states.}
  Plotted are equations of isotropically jammed  ($\gamma = 0$, solid lines) and  steady states ($\gamma = \infty$, dashed lines), where the (A) pressure $P$, (B) excess coordination number $\Delta Z$, and (C) macroscopic friction coefficient $\mu$ are plotted as  functions of volume fraction $\varphi$.  The vertical arrows present the constant-volume shear protocol.
  }
\label{fig:EOS}
\end{figure*}

\subsection*{ Shear hardening as a consequence of  memory loss by shear} To understand the origin of shear hardening, it is useful to examine the evolution of state on the generalized jamming phase diagram~\cite{Babu2021}. The key point is that, for a general $\varphi_{\rm j}$, the isotropic unstrained packings and the asymptotic stationary states at large strains satisfy different sets of equations of states (EOSs); they coincide only when $\varphi_{\rm j} \to \varphi_{\rm J}$ as in the initial proposal of the phase diagram by Liu and Nagel~\cite{Liu1998}.
The pressure, excess coordination number and macroscopic friction of unstrained states are described  by the following EOSs respectively, $P_{\rm 0} \sim \varphi - \varphi_{\rm j}$, $\Delta Z_{0} \sim (\varphi - \varphi_{\rm j})^{1/2}$, and $\mu_0 = 0$. The corresponding EOSs of stationary states are $P_{\rm s} \sim \varphi - \varphi_{\rm J}$,  $\Delta Z_{\rm s} \sim (\varphi - \varphi_{\rm J})^{1/2}$,  and  $\mu_{\rm c} -\mu_{\rm s} \sim (\varphi - \varphi_{\rm J})^{1/2}$, where  $\mu_{\rm c} \approx 0.1$~\cite{Peyneau2008, Zheng2018, Babu2021}.

Under constant volume shear, it is easy to see from Fig.~\ref{fig:EOS} that the system has to gain rigidity, because both the pressure and the number of contacts increase.
{The increase of anisotropy $\mu$ makes an additional contribution to the shear modulus, as demonstrated below.}
Shear hardening is thus a natural consequence of the memory erasing process caused by shear. Different unstrained states must evolve to the same stationary state that is independent of the initial condition. Initial states that are {dense} packings ($\varphi_{\rm j} \gg \varphi_{\rm J}$) created by deep annealing should also be brought back by shear to the generic packings at $\varphi_{\rm J}$ (corresponding to the rapid quench limit~\cite{OHern2003}) --  this process is accompanied by the increase of pressure and shear modulus under the constant volume condition. 


\subsection*{Microscopic origins of shear hardening} 
\begin{figure*}[!htb]
  \centering
  \includegraphics[width=0.9\linewidth]{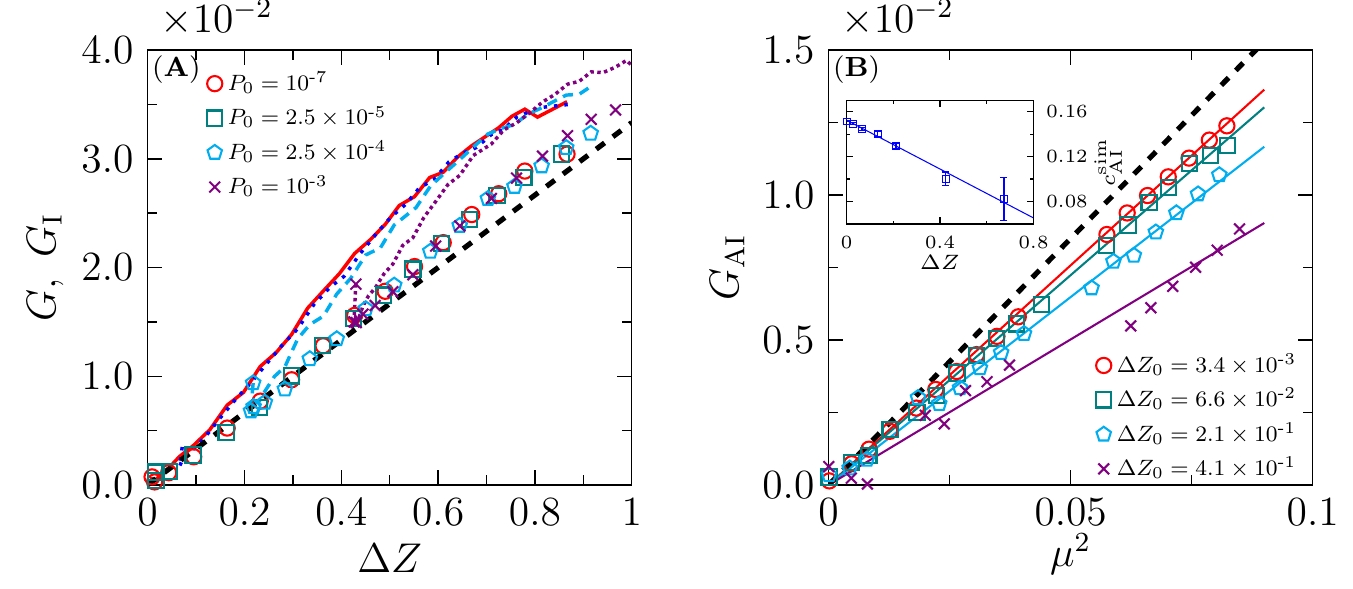}
  \caption{
  {Decomposition of the shear modulus into isotropic and anisotropic parts.}
  (A) Total shear modulus $G$ (lines) and its isotropic part $G_{\rm I}$ (points)
  as functions of $\Delta Z$.
  The {bold} dashed line corresponds to the theoretical result $G_{\rm I} = c_{\rm I} \Delta Z$ with $c_{\rm I} =  1/30$~\cite{Zaccone2011, Zaccone2011a}.
  {(B) Anisotropic part of the shear modulus $G_{\rm AI}$  as a function of $\mu^2$ for a few fixed $\Delta Z  = \Delta Z_0$ (see the legend in (A) for corresponding $P_0$ values).
  The {bold} dashed line indicts our theoretical result $G_{\rm AI} = c_{\rm 0} \mu^2$ {for $\Delta Z \to 0$,} with estimated  $c_{\rm 0} \approx 0.17$.
  The solid lines represent linear fitting to the data. The slopes are  plotted in the inset,  which are fitted to a linear function according to Eq.~(\ref{eq:cAI}), giving $c_{\rm AI}^{\rm sim}(\Delta Z) = 0.15 - 0.11 \Delta Z$ (line).}
  }
\label{fig:decomposition}
\end{figure*}

Previous studies have developed an elasticity theory for athermal disordered solids near the jamming transition, attributing compression hardening to the increase of interaction bonds via the relation~\cite{Wyart2005, Zaccone2011, Zaccone2011a},
\beq
G_{\rm I}  = c_{\rm I} \Delta Z,
\label{eq:GI}
\eeq
where $c_{\rm I} = 1/30$ (we have set both the bond stiffness $\kappa$ and the unstressed bond length $r_0=D$  to one).
While this theoretical result agrees well with the simulation data of sphere packings under isotropic compression~\cite{Zaccone2011a}, it cannot fully account for our shear hardening data (see Fig.~\ref{fig:decomposition}A).

\begin{figure}[!htb]
  \centering
  \includegraphics[width=\linewidth]{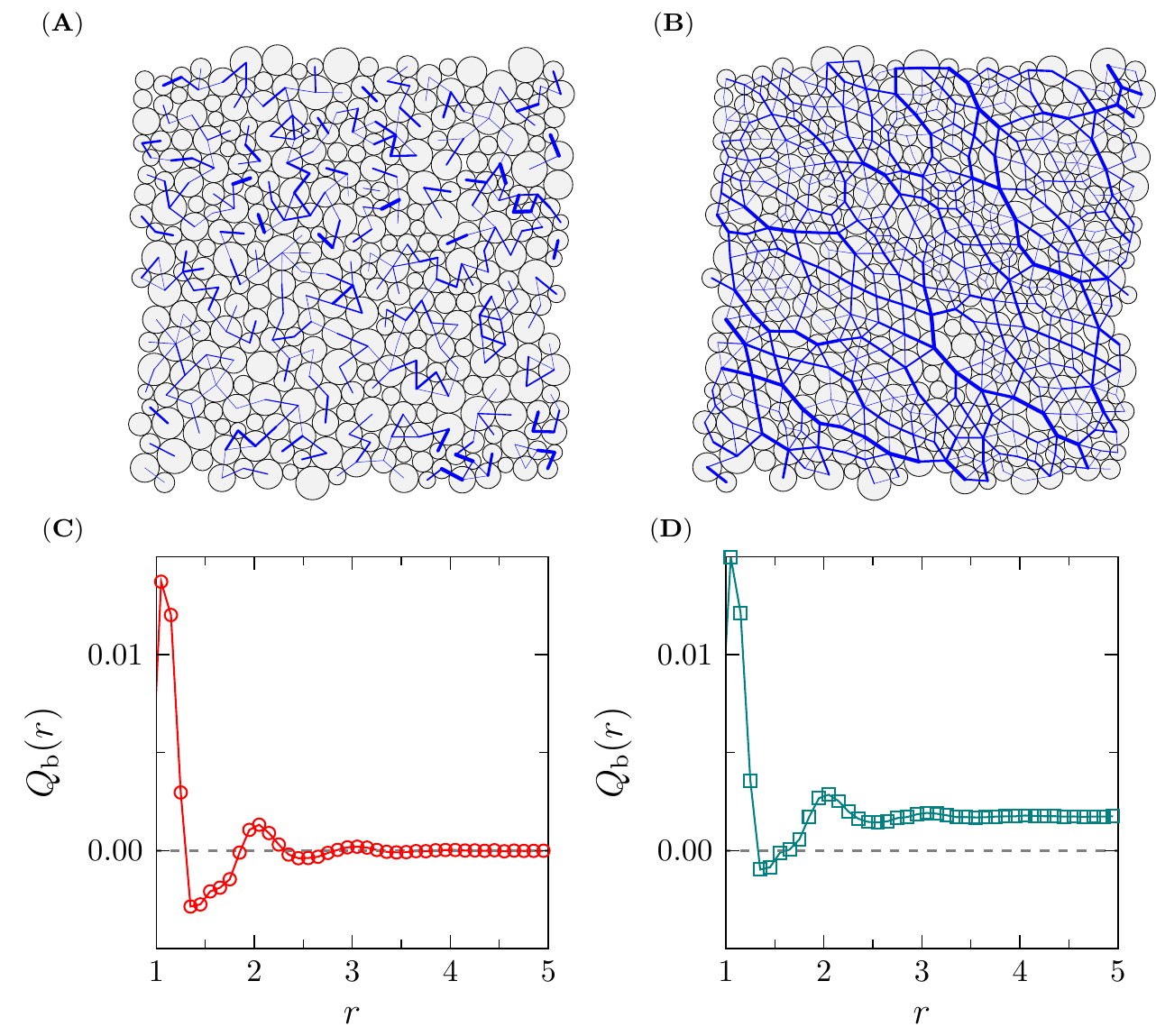}
  \caption{{Anisotropy and long-range orientational correlations of contact bonds.}
  (A-B) Typical  force networks of (A) isotropically compressed {($P_0 = 10^{-6},~\mu = 0$) and (B)  sheared samples
  ($P_0 = 10^{-6},~\mu = 0.36$)} in two dimensions, where the width of bond is proportional to the magnitude of contact force.
  (C-D) Bond orientation correlation function $Q_{\rm b}(r)$ of (C) isotropically compressed  ($P_0 = 10^{-7},~\mu = 0$) and (D)  sheared ($P_0 = 10^{-7},~\mu = 0.29$) 3D packings: $Q_{\rm b}(r \to \infty) = 0$  in (C) and $Q_{\rm b}(r \to \infty) > 0$ in (D).
  }
  \label{fig:bondCorr}
\end{figure}

To understand the microscopic origin of this deviation, we consider a generalized formula~\cite{Wyart2005} (see SI for details),
\beq
G = G_{\rm I} + G_{\rm AI},
\label{eq:G}
\eeq
{for $P \to 0$, where} $G_{\rm AI} \approx \frac{1}{N_{
\rm NR}} \sum_{ b_1 \neq b_2}^{N_{\rm b}} \tilde{f}_{b_1} \tilde{f}_{b_2} n_{b_1}^{x}  n_{ b_1}^{z}n_{ b_2}^{x}  n_{ b_2}^{z}$
depends on the normalized forces $\tilde{f}_b$ and bond orientations (represented by unit vectors $\vec{n}_b$), {$N_{\rm b}$ is the number of contacts and $N_{\rm NR}$ is the number of non-rattler particles (rattlers have fewer than $d+1$ contacts in $d$ dimensions)}. In isotropic packings, the lack of long-range spatial correlations between bond orientations, $\langle n_{b1}^x n_{b1}^z n_{b2}^x  n_{b2}^z \rangle \approx \langle n_{b1}^x n_{b1}^z \rangle \langle n_{b2}^x  n_{b2}^z \rangle =0$, suggests that $G_{\rm AI} \approx 0$ (note that $G_{\rm AI} \approx \langle \tilde{f}_{b1}\tilde{f}_{b2} \rangle\langle n_{b1}^x n_{b1}^z n_{b2}^x  n_{b2}^z \rangle $ under the effective media  approximation~\cite{Zaccone2011a}, and all contact forces $\tilde{f}_{b}$ are positive). In sheared packings, the bonds tend to align in shear-preferred directions and  are orientationally long-range correlated~\cite{BaityJesi2017}:  the correlation function,
$
Q_{\rm b} (r ) = \left \langle  n_{0}^x n_{0}^z n_{b}^x  n_{b}^z \delta \left(|\vec{r}_0 - \vec{r}_b| - r\right) \right \rangle,
$
where $\vec{r}_b$ is the position of bond $b$,
decays to  a finite value at large $r$ (see Fig.~\ref{fig:bondCorr}). Consequently,  $G_{\rm AI}$ is non-zero in such a case.

Our theoretical analysis predicts a simple relation between $G_{\rm AI}$ and the macroscopic friction $\mu$,
{
\beq
G_{\rm AI} = c_{\rm AI} (\Delta Z) \, \mu^2,
\label{eq:GAI}
\eeq 
where
\beq
c_{\rm AI} (\Delta Z) = c_0 - \alpha  \Delta Z,
\label{eq:cAI}
\eeq
with two constants $c_0$ and $\alpha$.
In the jamming limit ($\Delta Z \approx 0$), the coefficient $c_0$ can be evaluated from the formula, $c_0 = \frac{\langle f \rangle^2 }{3 \langle f^2 \rangle}$, where $\langle f \rangle$ and $\langle f^2 \rangle$ are the first two moments of the force distribution. In our system, we obtain from simulations $\frac{\langle f \rangle^2 }{\langle f^2 \rangle} \approx 0.5$, which gives $c_0 \approx 0.17$ (see Fig.~S6). Above jamming ($\Delta Z > 0$), $G_{\rm AI}$ is lowered by a higher-order correction term   $\sim \Delta Z \mu^2$.
}


To examine Eq.~({\ref{eq:GAI}}), we separate isotropic and anisotropic moduli in our simulation data as follows. Considering that $\Delta Z(\gamma)$ is increased during shear, at each $\gamma$ we first decompress the configurations keeping $\gamma$ unchanged, until $\Delta Z$ reaches $\Delta Z_0 \equiv \Delta Z(\gamma = 0)$ that is fixed by the initial condition $\Delta Z_0 \sim P_0^{1/2}$.
The decompressed configurations at different $\gamma$ thus now have the same $\Delta Z = \Delta Z_0$, but different $\mu$ {(we denote these configurations as a {\it constant-$Z$ ensemble})}. In this way, we remove the contributions from the change of $\Delta Z$, left with $G_{\rm AI}$ that changes with $\mu$.
Figure~\ref{fig:decomposition}B shows that $G_{\rm AI}$ is proportional to $\mu^2$ for any constant $\Delta Z$, as predicted by Eq.~(\ref{eq:GAI}). The behavior of the slope 
$c_{\rm AI}^{\rm sim}(\Delta Z)$ obtained from fitting the simulation data
is also consistent with our theoretical prediction Eq.~(\ref{eq:cAI}). 
The remaining shear modulus $G_{\rm I} =  G - G_{\rm AI}$ linearly depends on $\Delta Z$
(see Fig.~\ref{fig:decomposition}A), with a $P_0$-independent slope in agreement with the analytical result $c_{\rm I} = 1/30$ as in the case of pure compression~\cite{Zaccone2011a}. Note that here $\Delta Z$ is increased by shear instead of compression. { We find in our simulations that the pressure $P$ does not depend on $\mu$, but solely on $\Delta Z$. Thus the constant-$Z$ ensemble considered here is in fact equivalent to a constant-$P$ ensemble.}

Our results show that, within the first-order approximation, the excess coordination number  and anisotropy make additive contributions to 
the shear modulus.
At a large strain, the contact network becomes strongly anisotropic, in which case the anisotropic correction to $G$ is important.
On the other hand, in the limit $\sigma \to 0$ (or $\gamma \to 0$), the anisotropic contribution is subleading:
combining the scalings $\Delta Z \sim \sigma^{2/5}$ and $\mu \sim \sigma ^{1/4}$ (Figs.~\ref{fig:results_stress}B-C) with Eqs.~(\ref{eq:GI},~\ref{eq:G},~\ref{eq:GAI}), one obtains $G \sim \Delta Z \sim \sigma^{2/5}$ (Fig.~\ref{fig:results_stress}A).

\begin{figure}
  \centering
  \includegraphics[width=0.9\linewidth]{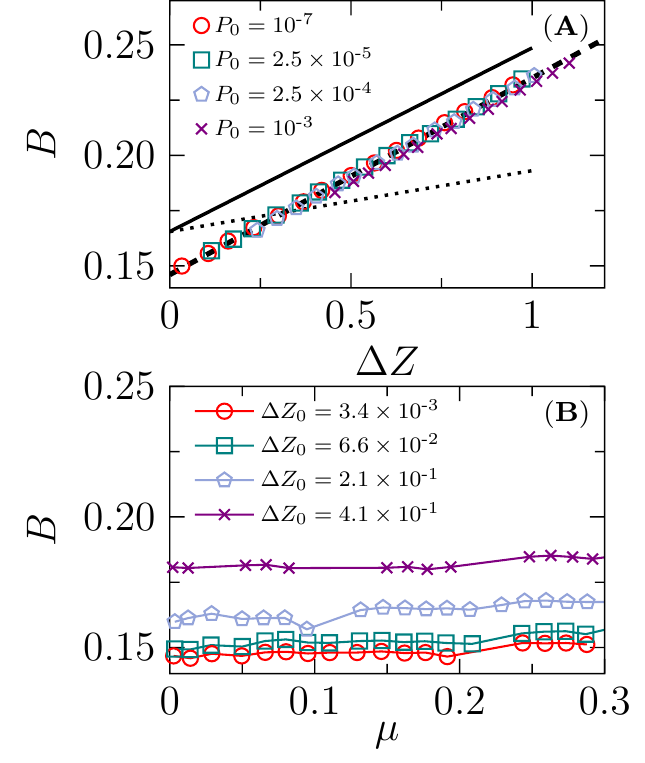}
  \caption{
  Bulk modulus of simulated systems. (A) Bulk modulus $B$ of strained configurations as a function of $\Delta Z$ for several $P_0$. The dashed line represents the linear fitting, $B = 0.147 + 0.0896 \Delta Z$. The solid and dotted lines are the Eq.~(\ref{eq:bulk}) and Reuss estimation with $c_0 \approx 0.17$. (B) Bulk modulus $B$ of samples with a varying $\mu$. Each curve has a constant   $\Delta Z  = \Delta Z_0$, which is realized by decompression as in Fig.~\ref{fig:decomposition} (see the legend in (A) for the corresponding $P_0$ values).
  }
\label{fig:bulk}
\end{figure}

For comparison, we also derive a theoretical expression of the bulk modulus $B$ near the jamming transition (see SI Sec. 6), 
\beq
B = c_0 + \left(\frac{1}{18}+\frac{c_0}{6} \right) \Delta Z,
\label{eq:bulk}
\eeq
where the constant  $c_0 = \frac{\langle f \rangle^2 }{3 \langle f^2 \rangle}$ is identical to the one appeared in Eq.~(\ref{eq:cAI}) for the  shear modulus. In the jamming limit $\Delta Z \to 0$,  the bulk modulus is a constant $ B(\Delta Z = 0) = c_0$.
In other words, $B$ changes discontinuously at either compression or shear jamming, following a scaling $B \sim \Delta Z^0$ as previously reported~\cite{OHern2003,BaityJesi2017}. Equation~(\ref{eq:bulk}) further gives the next order correction that depends linearly on $\Delta Z$. Plugging $c_0 \approx 0.17$ into Eq.~(\ref{eq:bulk}) gives $B \approx 0.17 + 0.084 \Delta Z$, which is close to the result  $B = 0.146 + 0.0897 \Delta Z$ obtained from fitting the simulation data (see Fig.~\ref{fig:bulk}A). As expected and suggested by Eq.~(\ref{eq:bulk}), the bulk modulus $B$ is  independent of the anisotropy $\mu$ (see Fig.~\ref{fig:bulk}B for the numerical verification).


In our theoretical analysis, the internal stresses is ignored.
Because the considered interaction is purely repulsive, the non-zero internal stress should lower the energy and therefore the modulus. In other words, the modulus obtained from the current theory overestimates the actual value. Approaching the jamming transition where the internal stress  vanishes,  the current theoretical results are very close to the simulation data, suggesting that the effects of internal stress are not significant in this limit (see Figs.~\ref{fig:decomposition} and ~\ref{fig:bulk}).

Interestingly,  a previous study~\cite{ peyneau2008solidlike} provided an approximated  expression 
$B^{\rm Reuss} = \frac{Z \langle f \rangle^2}{18 \langle f^2 \rangle} = \frac{Z c_0}{6}$ for the bulk modulus  using the so-called Reuss estimate~\cite{agnolin2007internalc}.
In Fig.~\ref{fig:bulk}, we compare $B^{\rm Reuss}$ with our theory Eq.~(\ref{eq:bulk}) and simulation data. 
To make a further comparison, we expand $B^{\rm Reuss}$ to the linear order of $\Delta Z$, using $Z \approx 6 + \Delta Z$, which results in
 $B_{\rm }^{\rm Reuss} \approx c_0 + \frac{c_0}{6} \Delta Z$.
Compared to our theoretical result Eq.~(\ref{eq:bulk}), this final expression gives the same constant $c_0$ when $\Delta Z \to 0$, but 
 a different
 coefficient of the linear term.

\color{black}
\section*{Discussion}

Our finding raises several challenges to the theoretical understanding of jamming transition and rheology of amorphous solids. In particular, the scaling {$\mu \sim \sigma^{\beta}$ with $\beta \approx 0.25$}, which is universal in two to four dimensions (see SI), requires {an quantitative} theoretical explanation.
The scaling ansatz proposed in Ref.~\cite{Goodrich2016}, $G=\Delta Z \mathcal{G}_0\left(\frac{\Delta \varphi}{\Delta Z^2}, \frac{\gamma}{\Delta Z^{3/2}}, N \Delta Z \right)$, treats $\Delta Z$ as the only relevant order parameter for the jamming transition. Equation~(\ref{eq:G}) suggests that the ansatz needs to be extended, since the anisotropic part $G_{\rm AI}$, which relies on  $\mu$ instead of $\Delta Z$, contributes independently to the total shear modulus.

The response of amorphous solids to compression and shear has been studied by a first-principle mean-field theory -- the state-following glass theory~\cite{Rainone2015, rainone2016following}. In the stable-glass phase, the theoretical outcome is $\sigma \sim G \gamma$ and $P \sim P_0 + R \gamma^2$, where $R$ is the dilatancy parameter (note that the second relation reflects the symmetry $P(\gamma) = P(-\gamma)$).  These relationships give a scaling $\mu  \sim \sigma$ in the limit $\gamma \to 0$, which can only explain the first scaling (in the elastic and softening regimes) in Fig.~\ref{fig:collapse}(B). The issue is twofold: on the one hand, it remains technically difficult to perform the state-following computation in the marginally stable phase where jammed states belong to~\cite{rainone2016following}; on the other hand, conceptually one should consider the possibility of "state rejuvenation" since simulations suggest $\varphi_{\rm j}$ decreasing towards  $\varphi_{\rm J}$ in the shear hardening regime~\cite{Kawasaki2020, pan2022nonlinear}.


Finally, it would be interesting to examine the effect of friction on shear hardening. Recent progress in producing ultra-stable granular materials makes an experimental investigation of frictional shear hardening feasible~\cite{zhao2022ultrastable, wang2022experimental}.
It is expected that the critical exponents are non-universal with different friction coefficients~\cite{wang2022experimental}.
In addition, the correction to the isotropic shear modulus should be even more significant, because friction amplifies spatial heterogeneity and orientational anisotropy of the contact network.

\subsection*{Simulation models} We consider a 3D granular model of frictionless poly-disperse spheres  with a continuous diameter distribution $P(D) \sim D^{-3}$, for $D_{\rm min} \leq D \leq D_{\rm min}/0.45$~\cite{Berthier2016, Ninarello2017, jin2021jamming}. The particles interact via a harmonic SS pair potential, $v_{ij}(r_{ij}) = \frac{1}{2} (1- r_{ij}/D_{ij})^2$ (zero if $r_{ij}>D_{ij}$), where $r_{ij}$ is the inter-particle distance and $D_{ij} = (D_i + D_j)/2$ the mean  diameter of particles $i$  and $j$.
{ We set the average diameter $\overline{D}$ as the unit of length.}
The lowest jamming density, or the J-point density~\cite{OHern2003}, of this model is $\varphi_{\rm J} \approx 0.655$~\cite{Ozawa2017,jin2021jamming}.
The reported results are obtained from systems of $N = 2000$ particles, averaged over $192$ independent samples.
In addition, we study 2D and 4D models with the same kind of interaction in SI.

\subsection*{Preparation of initial states} Two approaches are used to prepare ultra-stable jammed configurations. (i) Swap thermal annealing. The thermal annealing is realized by applying an efficient swap Monte Carlo algorithm~\cite{Berthier2016, Ninarello2017}, which generates equilibrium hard-sphere (HS) configurations at $\varphi_{\rm eq}$ above the mode-coupling theory (MCT) crossover density $\varphi_{\rm MCT} \approx 0.594$~\cite{Berthier2016a}. After that, we set the temperature to zero in the remaining simulations,
and switch to  the SS potential $v_{ij}(r_{ij})$.
By applying athermal
quasi-static compressions (AQCs) with the FIRE energy minimization algorithm~\cite{Bitzek2006} as described in~\cite{Chaudhuri2010,Ozawa2017,Babu2021},  we obtain jammed configurations at $\varphi_{\rm eq}$-dependent jamming densities
$\varphi_{\rm j}$~\cite{Ozawa2012a,Ozawa2017,jin2021jamming}.
In the main text, we focus on the case of $\varphi_{\rm j}=0.69$ unless otherwise specified, corresponding to $\varphi_{\rm eq} = 0.643$. (ii) Mechanical training. The method
is realized by cyclic AQS~\cite{Babu2021, Kawasaki2020, Das2020, otsuki2020shear} or cyclic AQC~\cite{Kumar2016}.
Random initial configurations are rapidly compressed over $\varphi_{\rm J}$, and become unjammed after
a sufficient number of cyclic AQS or AQC under the constant volume condition~\cite{Babu2021}, meaning that the jamming density is increased to $\varphi_{\rm j} > \varphi_{\rm J}$. In both approaches,  $\varphi_{\rm j}$ correlates to the stability of the jammed states, and reflects the degree of annealing.

The above procedures generate packings at $\varphi = \varphi_{\rm j}$ and $P = 6\times10^{-9}$, which are extremely close to the jamming/unjamming transition.
They are then compressed quasi-statically to over-jammed states at a series of pressures $P_0 \equiv P(\gamma = 0)$. The pressure $P_0$ characterizes the distance
to the jamming point, via the well-known scaling between  $P_0$ and the  density $\varphi_0$ for harmonic SSs~\cite{OHern2003}, $P_0 \sim (\varphi_0 - \varphi_{\rm j})$.
In order to remove residual stresses, the shear stabilization method~\cite{DagoisBohy2012} is applied, and these configurations with zero residual stresses are referred to as the {\it unstrained states} {($\sigma = 0$).
According to this definition, in finite-size unstrained states, the stress $\sigma$ is always zero, while the strain $\gamma$ can fluctuate around zero from sample to sample. 
In our scaling analysis (see Fig.~\ref{fig:results_stress}), it is better to use  $\sigma$   than  $\gamma$ as the independent variable, since the latter can introduce additional uncertainties. 
Of course, in the limit of large sample size and/or of larger number of samples, this sample-to-sample fluctuation disappears. In fact,  a similar issue exists in the case of compression jamming: the jamming transition is defined at the zero pressure, while the jamming density can 
 fluctuate around the mean value  in finite-size simulations. Thus practically it is better to use the pressure than the density as the independent variable in  scaling analyses.}

\subsection*{Compression and shear protocols} { Compression and decompression procedures are realized by rescaling particle diameters with energy minimization.}
Starting from the unstrained states,  we apply constant-volume, simple AQS in the  $x$-$z$ plane~\cite{Babu2021,jin2021jamming}.
At each step, the shear strain $\gamma$ is increased by $\delta \gamma$, followed by the FIRE energy minimization, where $\delta \gamma$  is logarithmically  increased from $10^{-8}$ to $2\times10^{-4}$.

The stress and pressure are calculated using the virial formula~\cite{tsai1979virial},
$\sigma = -\frac{1}{N_{\rm NR}}\sum_{b=1}^{N_{\rm b}} r_{b}^x f_{b}^z $
and $P = \frac{1}{3N_{\rm NR}}\sum_{b=1}^{N_{\rm b}} \vec{r}_{b}  \cdot  \vec{f}_{b} $,
where $\vec{r}_{b}$ and $\vec{f}_{b}$ are the contact vector and force on bond $b$,
$N_{\rm NR}$ the number of non-rattler particles (rattlers have fewer than $d+1$ contacts in $d$ dimensions), and $N_{\rm b}$ the total number of bonds. The shear modulus is defined as $ G = \langle \delta \sigma/\delta \gamma \rangle + \langle P \rangle $, averaged over 192 independent samples, where $\delta \gamma$ is between $10^{-10}$ and $10^{-8}$. 
{ The term  $\delta \sigma/\delta \gamma$ corresponds to the slope of a single-realization stress-strain curve, which is piecewise linear~\cite{dubey2016elasticity} (see the inset of Fig. 1B).}
The correction term $\langle P \rangle$, which is typically three orders smaller than the first term, is added so that $G$ is equivalent to the $xzxz$ component of the stiffness matrix. We have checked that the value of $G$ obtained in this way is identical to that directly calculated using the explicit expressions derived from the elastic theory~\cite{Karmakar2010athermal}. {The bulk modulus is computed by the formula $B = -\langle \delta P /\delta \epsilon_{\rm V}\rangle -\frac{1}{3} \langle P \rangle$, where $\epsilon_{\rm V}$ is the volumetric strain and $ \delta \epsilon_{\rm V} \leq 2\times10^{-9}$.}

The changes of two independent parameters $Z$ and $\mu$ reflect local rearrangements of particles.
The increase in the number of  contacts
is quantified by the excess coordination  number, $\Delta Z = Z - Z_{\rm iso}$,
where  $Z_{\rm iso} = 2d - 2d/N_{\rm NR}$
is the coordination number required to satisfy the isostatic condition at the jamming transition in $d$ dimensions with a finite size correction~\cite{Goodrich2012, Goodrich2014},
{ and rattlers are excluded in  the computation of $Z$.}
The macroscopic friction coefficient $\mu = \sigma/P$ measures the degree of anisotropy caused by shear~\cite{Chen2018}.

\subsection*{Data availability} All data are available in the main text or the supplementary information.




\subsection*{Acknowledgements} We warmly thank Hajime Yoshino and Eric DeGiuli for discussions.
This work was supported by
funding from the National Natural Science Foundation of China (Project 12161141007, Projects 11974361, Project 11935002, and Project 12047503), and the Chinese Academy of Sciences (the Key Research Program of Frontier Sciences Grant NO. ZDBS-LY-7017, the Key Research Program Grant NO. XDPB15, Grant NO. KGFZD-145-22-13, and Grant NO. XDA17010504). In this work access was granted to the High-Performance Computing Cluster of Institute of Theoretical Physics the Chinese Academy of Sciences. This manuscript was posted on a preprint: https://doi.org/10.48550/arXiv.2208.08793.

\clearpage

\centerline{\bf Supplementary Information}

\setcounter{figure}{0}
\setcounter{equation}{0}
\setcounter{table}{0}
\renewcommand\thefigure{S\arabic{figure}}
\renewcommand\theequation{S\arabic{equation}}
\renewcommand\thesection{S\arabic{section}}
\renewcommand\thetable{S\arabic{table}}

\section{Shear strain as the independent variable}

\begin{figure*}
  \centering
  \includegraphics[width = 0.9\linewidth]{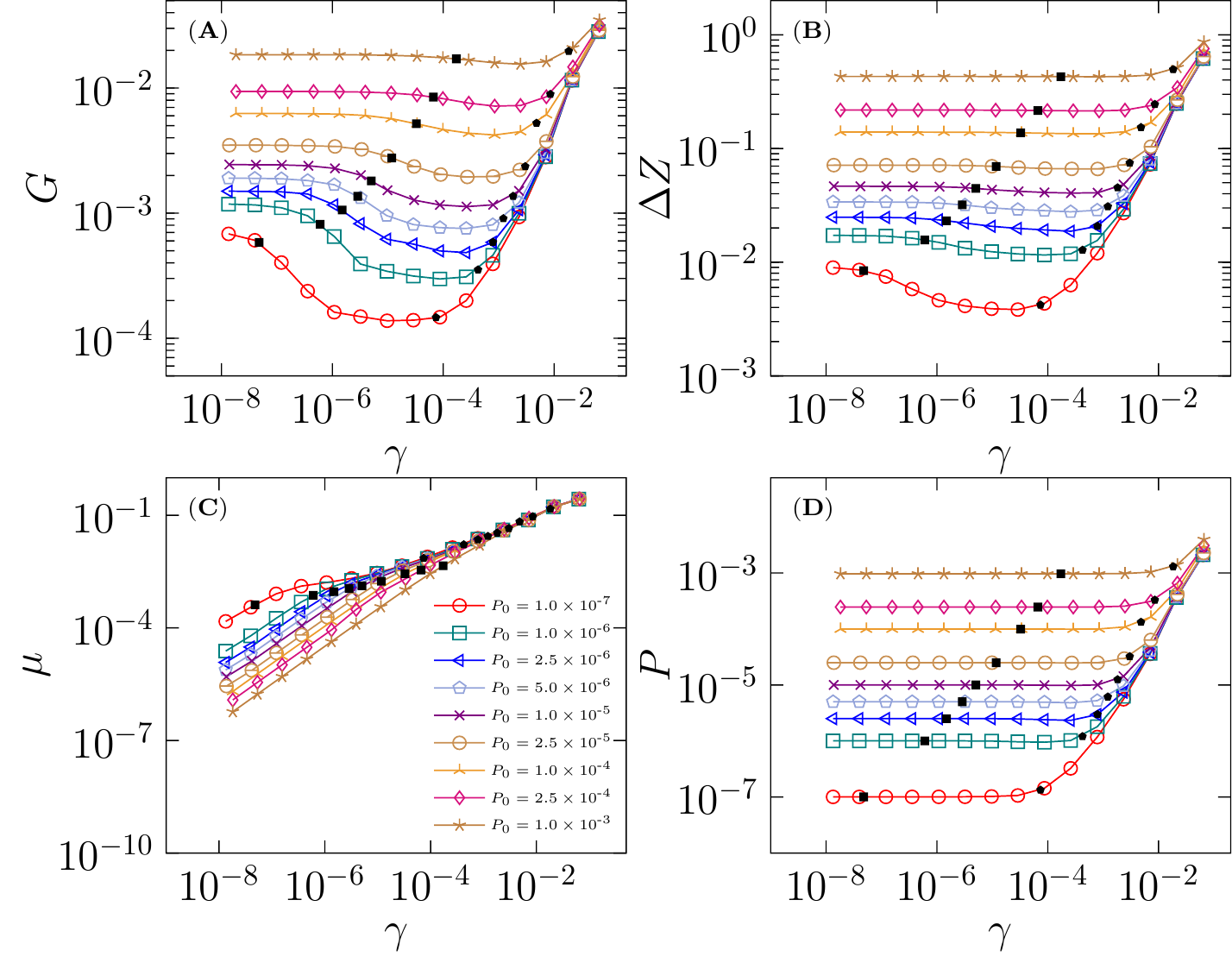}
  \caption{{
  Simulation results of (A) the shear modulus $G$, (B) the excess coordination number $\Delta Z$, (C) the macroscopic friction coefficient $\mu$ and (D) the pressure $P$ as functions of strain $\gamma$. The crossovers $\gamma_{\rm s}$ and $\gamma_{\rm h}$ are represented by solid squares and pentagons, respectively. }
  }
  \label{fig:strain_loglog}
\end{figure*}

In our simulations, the stress $\sigma$ has much smaller fluctuations than  the strain $\gamma$ (see {\it Materials and Methods}). Thus $\sigma$ is chosen as the independent variable in our scaling analyses (see Fig.~3). For completeness, in Fig.~\ref{fig:strain_loglog}, we plot $G$, $\Delta Z$, $\mu$ and $P$ as functions of $\gamma$. The data are essentially equivalent to those presented in Fig.~3, with $\gamma$ treated as the independent variable.

\section{Independence of scaling exponents on the degree of annealing}

\begin{figure*}
  \centering
  \includegraphics[width = 0.9 \linewidth]{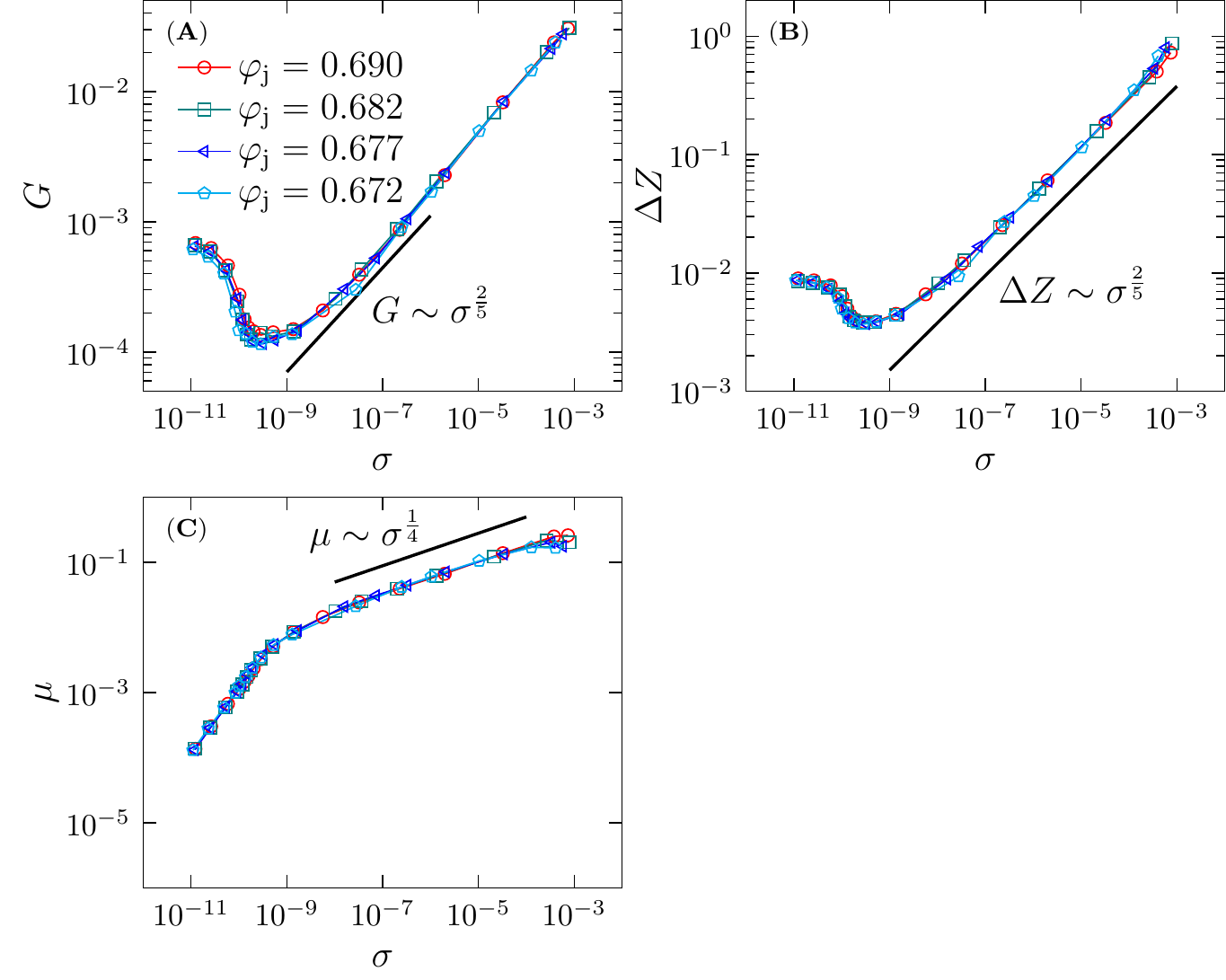}
  \caption{{Shear hardening scalings of systems with different degree of annealing.} Simulation results of (A) shear modulus $G$, (B) excess coordination number $\Delta Z$ and (C) macroscopic friction coefficient $\mu$, as functions of stress $\sigma$, for a few different $\varphi_{\rm j}$. The unstrained pressures are the same,  $P_0 = 10^{-7}$. The solid lines represent scaling laws in the shear hardening regime, $G\sim \sigma^{2/5}$, $\Delta Z \sim \sigma^{2/5}$ and $\mu \sim \sigma^{1/4}$.}
  \label{fig:results_stress_phig}
\end{figure*}

In the main text, we show that shear hardening presents in deeply annealed packings ($\varphi_{\rm j} = 0.69$).
Here, we perform additional simulations on systems with different  $\varphi_{\rm j}$ (i.e., different degrees of annealing).
The data in Fig.~\ref{fig:results_stress_phig} show that the exponents in all three scalings, $G \sim \sigma^{2/5}$, $\Delta Z \sim \sigma^{2/5}$ and $\mu \sim \sigma^{1/4}$, are independent of $\varphi_{\rm j}$. Although the curves of different $\varphi_{\rm j}$ seem to collapse in the figure completely, one should note that the yielding stress vanishes ($\sigma_{\rm Y} \to 0$) in the limit of $\varphi_{\rm j} \to \varphi_{\rm J}$~\cite{pan2022nonlinear}, and consequently so does the scaling regime of shear hardening.

\section{Shear hardening in systems prepared by mechanical training}

Dense packings with a large $\varphi_{\rm j}$ can be generated by either thermal or mechanical annealing. The latter refers to athermal training procedures such as cyclic shear or cyclic compression.
The simulation data presented in the main text are obtained from
well-annealed systems prepared by swap thermal annealing. Here, we present the results of samples annealed by mechanical training, and show that the scaling exponents are independent of the annealing protocol.

Two different training methods, cyclic shear and cyclic compression, are used.
In the cyclic shear protocol~\cite{Babu2021}, random configurations are firstly over-compressed to a density $\varphi_x > \varphi_{\rm J}$. Then, multiple cycles  of shear, $0 \rightarrow \gamma_{\rm max} \rightarrow - \gamma_{\rm max} \rightarrow 0$, are applied until the system becomes unjammed at $\varphi_x$.
In the cyclic compression protocol~\cite{Kumar2016}, the random configurations are over-compressed to a pressure $P_{\rm max}$, and then are decompressed to $\varphi = 0.62$ that is lower than the jamming density. Such cycles are stopped when the desired jamming density $\varphi_{\rm j} = 0.663$ is reached.
Both mechanical training procedures are performed under the athermal quasi-static condition.
The strain increment is $\delta \gamma = \pm 10^{-3}$ and the density increment is $\delta \varphi = \pm 10^{-4}$.

The jamming density  $\varphi_{\rm j}$ depends on the parameters $\gamma_{\rm max}$ and $\varphi_{x}$ in the cyclic shear protocol, and
$P_{\rm max}$  in the cyclic compression protocol~\cite{Babu2021, Kumar2016}. In this study, we set $\gamma_{\rm max} = 0.05$ and $\varphi_{x} = 0.66$ in the former, and $P_{\rm max} = 10^{-2}$  in the latter.
With these parameters, we obtain $\varphi_{\rm j} = 0.662$ and $\varphi_{\rm j} = 0.663$ respectively, both of which are above the J-point density $\varphi_{\rm J} \approx 0.655$.

\begin{figure*}
  \centering
  \includegraphics[width= 0.9 \linewidth]{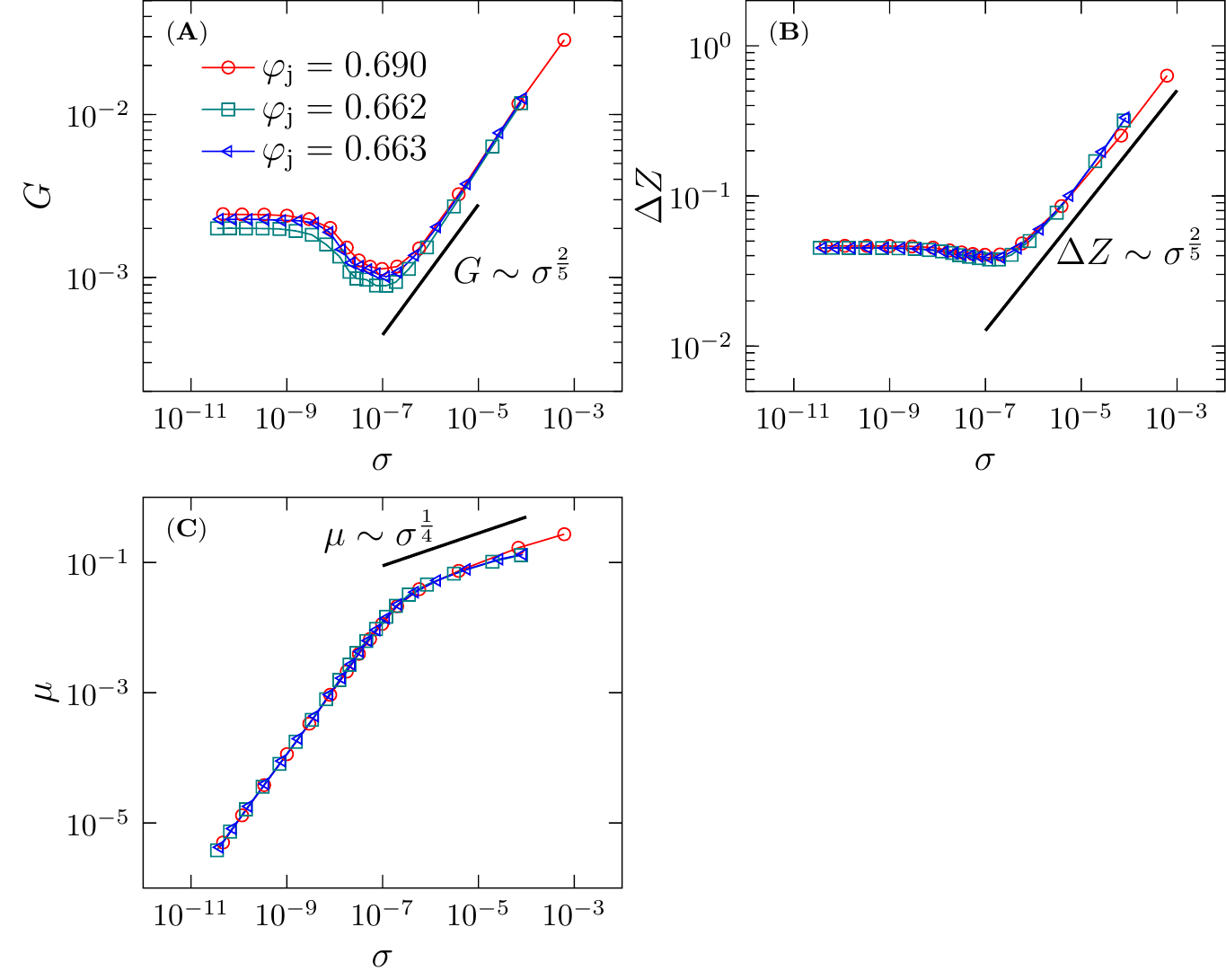}
  \caption{{Shear hardening scalings of systems prepared by thermal annealing and mechanical training.} Simulation results of (A) shear modulus $G$, (B) excess coordination number $\Delta Z$ and (C) macroscopic friction coefficient $\mu$, as  functions of stress $\sigma$, for systems prepared by swap thermal annealing ($\varphi_{\rm j} = 0.69$), cyclic shear ($\varphi_{\rm j} = 0.662$) and cyclic compression ($\varphi_{\rm j} = 0.663$). The unstrained pressures are the same,  $P_0 = 10^{-5}$.}
  \label{fig:annealProtocol}
\end{figure*}

These mechanically trained systems also show shear hardening behavior (Fig.~\ref{fig:annealProtocol}), with scalings identical to those in Fig.~3, which are obtained from systems generated by swap thermal annealing. We thus conclude that the scaling laws of shear hardening are independent of the way of annealing.

\section{Shear hardening in two and four dimensions}

The systems are composed by $N = 4000$ (2D) and $N = 2000$ (4D) particles. The diameter distribution is $p(D) \sim D^{-d}$, for $D_{\rm min} \le D \le D_{\rm max}/0.45$. As in the 3D model, two particles interact with each other via a harmonic soft sphere potential $v_{ij}(r_{ij})$, if they are in contact.
The mean diameter of all particles is set as the unit length. All particles have the same unit mass.

In 2D, we use the swap thermal annealing protocol to prepare equilibrium hard disk liquids of density $\varphi_{\rm eq} = 0.85$. After that, we switch to the soft potential, and compress the system to a jammed state of $\varphi_{\rm j} = 0.881$ (the J-point density is $\varphi_{\rm J}  = 0.841$).

We find that it is very easy to observe shear hardening in 4D.
Deep annealing is unnecessary-- it is enough to perform only one  compression-decompression cycle. The jamming density  of the obtained samples is $\varphi_{\rm j} = 0.486$, which is only slightly larger than the the J-point density $\varphi_{\rm J} = 0.480$.

\begin{figure*}
  \centering
  \includegraphics[width=0.9\linewidth]{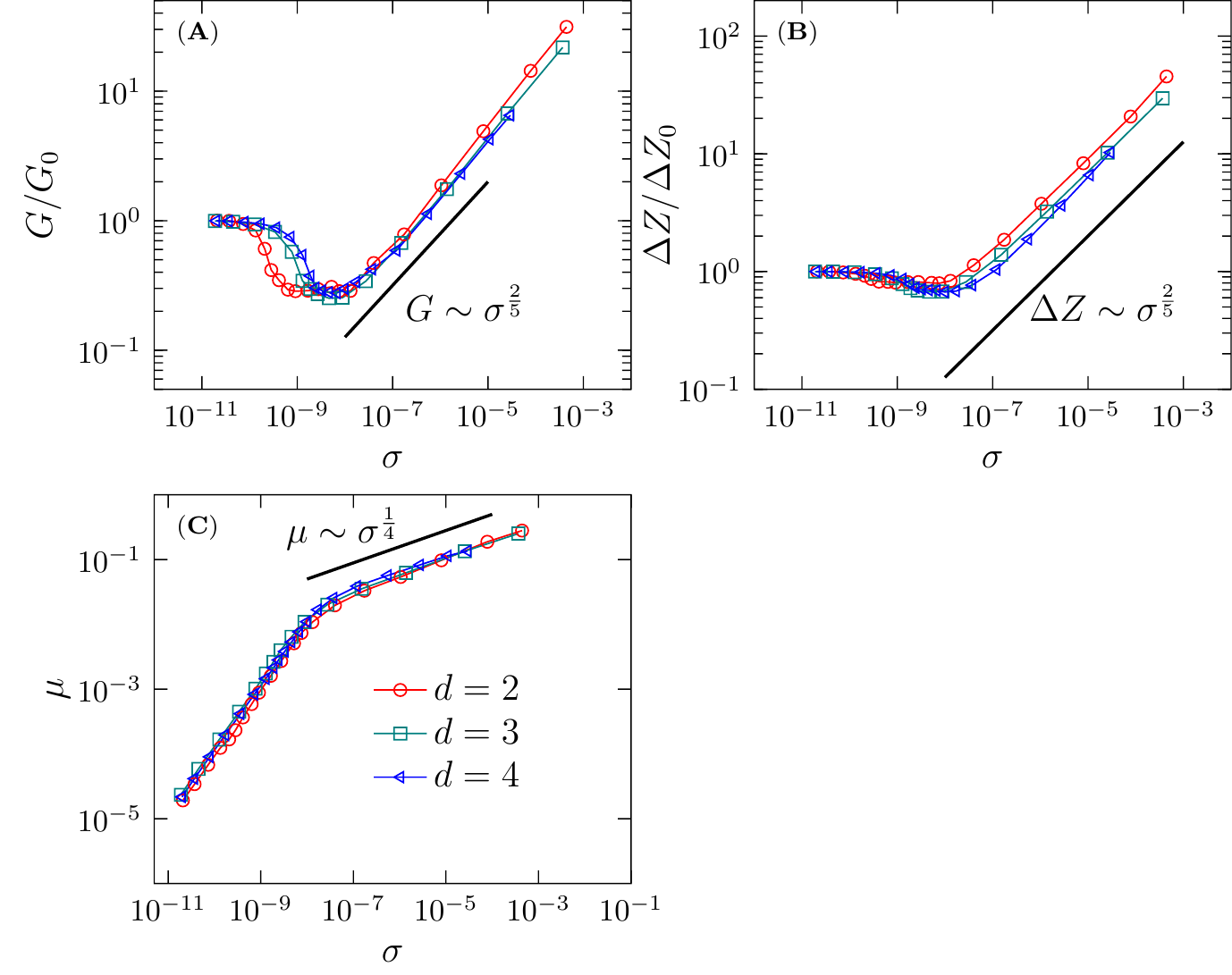}
  \caption{{Shear hardening scalings of systems in two to four dimensions.} Simulation results of (A) shear modulus $G$, (B) excess coordination number $\Delta Z$ and (C) macroscopic friction coefficient $\mu$ versus stress $\sigma$ in 2D ($\varphi_{\rm j} = 0.881$), 3D ($\varphi_{\rm j} = 0.690$) and 4D($\varphi_{\rm j} = 0.486$), for $P_0 = 10^{-6}$.
  In (A) and (B), $G$ and $\Delta Z$ are scaled by their unstrained values $G_0$ and $\Delta Z_0$ respectively.
  }
  \label{fig:dimScaling}
\end{figure*}

The scaling behavior in 2D and 4D  is reported in Fig.~\ref{fig:dimScaling}, together with the data obtained in 3D.
The scaling exponents in the shear hardening regime are nearly independent of the dimensionality $d$. This observation is consistent with the previous conjecture of upper critical dimension $d_{\rm u} = 2$ for the jamming transition~\cite{Liu2010, Goodrich2012}.

\section{Finite-size effects}

\begin{figure*}
  \centering
  \includegraphics[width = 0.9 \textwidth]{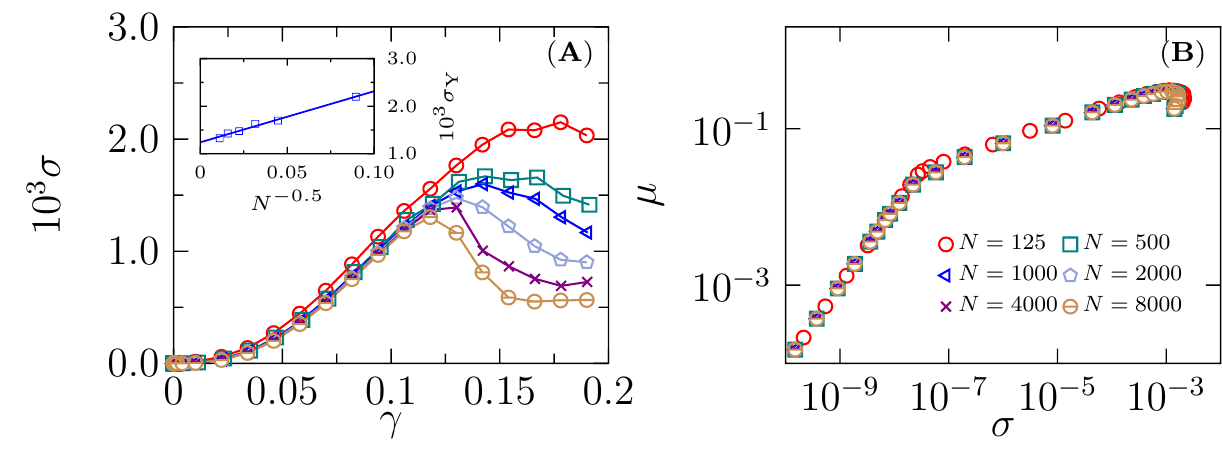}
  \caption{{
  (A) Stress-strain curves and (B) macroscopic friction coefficient $\mu$ as a function of stress $\sigma$ for different system sizes $N$, with fixed  $P_0 = 10^{-6}$ and $\varphi_{\rm j} = 0.69$.  The system size dependence of yielding stress $\sigma_{\rm Y}$ (i.e., the maximum stress) is shown in the inset of (A) and the solid line is the fitting curve $\sigma_{\rm Y}(N) = 0.0107  N^{-0.5} + 0.00124$.
  } 
  }
  \label{fig:finiteN}
\end{figure*}

To check finite size effects, we consider systems of different $N$, with
the jamming density $\varphi_{\rm j} = 0.69$  and the unstrained initial pressure $P_0 = 10^{-6}$  fixed. The stress-strain curves  with different  $N$ nearly coincide with each other before yielding (see Fig.~\ref{fig:finiteN}A). The yielding stress $\sigma_{\rm Y}$ dependents on $N$, satisfying  $\sigma_{\rm Y}(N) - \sigma_{\rm Y}(N = \infty) \sim N^{-0.5}$ (see the inset of Fig.~\ref{fig:finiteN}A). In Fig.~\ref{fig:finiteN}B, we plot  $\mu$ as a function of $\sigma$, and find that there is no noticeable finite size effects on the crossover stress $\sigma_{\rm h}$. 
Thus the shear-hardening regime $\sigma_{\rm  h} < \sigma < \sigma_{\rm Y}$ becomes smaller in larger systems, but remains finite in the thermodynamic limit.

\section{Elasticity theory}

\subsection{Setup}

{\it Model.} The model consists of $N$ soft spheres {(rattlers are not considered in this theoretical model)} interacting via a harmonic potential at zero temperature. The total energy is
\beq
E  = \sum_{i < j}  v_{ij}(r_{ij}),
\label{eq:E}
\eeq
where (we set both the spring constant and  particle diameter to be one)
\beq
v_{ij}(r_{ij}) = \frac{1}{2}(1-r_{ij})^2.
\label{eq:vij}
\eeq
The vector $\vec{r}_{ij} = \vec{r}_j - \vec{r}_i$ goes from particle $i$ to $j$. We are interested in the response of the model to athermal quasi-static shear, meaning that the system is force balanced at each shear step. Without loss of generality, we only compute the $xzxz$ component  {$G = \mathcal{C}^{xz, xz}$} of the stiffness tensor $\mathcal{C}$
as the shear modulus.

The following approximations are used, which are non-essential for the problem under consideration.
(i) In simulations, the interaction is purely repulsive, and therefore Eq.~(\ref{eq:vij}) is truncated for $r_{ij} > 1$.
The theoretical potential is perfectly symmetric. Thus the amorphous solid is modeled by a spring network.
(ii) The polydispersity is neglected. (iii) The pre-stress is ignored in the calculation of shear modulus.

{\it Notations.} To avoid confusion, we use letters with arrows for spatial vectors in three
dimensions (e.g.,  $\vec{r}_i = \{r_i^x, r_i^y, r_i^z\}$), and bold letters for general vectors  (e.g., ${\bf v}, {\bf \Xi}, {\bf f}, \ldots$).
Roman letters ($i, j, \ldots$) denote particle indices, and Greek characters ($\alpha, \beta, \ldots$) denote cartesian components.

{\it Angular averages.} The orientation of any interaction bond can be denoted by a unit vector in three dimensions, $\vec{n} = \{ \sin(\theta)\cos(\phi),\cos(\theta),\sin(\theta)\sin(\phi) \}$.  For a simple shear deformation in the $x$-$z$ plane, the  bond angles are distributed according to a function,
\beq
\rho(\theta,\phi) = \frac{1}{4\pi} - \frac{R_{\rm A}}{2\pi}\sin(2\phi),
\label{eq:rho}
\eeq
characterized by the fabric anisotropy parameter $R_{\rm A}$ that can be computed from the fabric tensor~\cite{Radjai1998, jin2021jamming}.
The fabric tensor is defined as,
\beq
\mathcal{R} = \frac{1}{N} \sum_{b=1}^{N_{\rm b}} \vec{n}_{b} \otimes  \vec{n}_{b},
\eeq
where $N_{\rm b}$ is the total number of contacts, and
$\otimes$ is a vector outer product. The coordination number $Z$ is the sum of eigenvalues of $\hat{R}$, $Z= \lambda_1+\lambda_2+\lambda_3$, and the fabric anisotropy parameter $R_{\rm A}$ is defined as  $R_{\rm A} = (\lambda_{\rm max} - \lambda_{\rm min})/Z$.

With Eq.~(\ref{eq:rho}), the average of any angular dependent quantity can be calculated using the formula,
\beq
\left \langle n^{\alpha} n^{\beta}  \ldots n^{\chi} \right \rangle  =    \int_0^{\pi} \sin \theta d\theta \int_0^{2 \pi} d \phi \, \rho(\theta, \phi) n^{\alpha} \, n^{\beta}  \ldots n^{\chi}.
\label{eq:ave}
\eeq
In particular,
\beq
\left \langle n^{x} n^{z}   \right \rangle = {-\frac{R_{\rm A}}{3}},
\eeq
and
\beq
\left \langle n^{x} n^{z} n^{x} n^{z}   \right \rangle = \frac{1}{15},
\label{eq:n4}
\eeq
which is independent of $R_{\rm A}$ and will be used below.

{\it Operators.} In Ref.~\cite{Wyart2005},
Wyart introduces two operators $\mathcal{S}$ and $\mathcal{T}$, which make conversions between particle-based vectors (size $3N$) and bond-based vectors (size $N_{\rm b}$).
We briefly review their definitions and properties.

The matrix $\mathcal{S}$ (a $N_{\rm b}$ by $3N$ matrix)  converts a particle-based displacement field ${\bf \delta R}$ (a vector of size $3N$) to displacements ${\bf \delta r}$  along interaction bonds (a vector of size $N_{\rm b}$):
\beq
\mathcal{S} {\bf \delta R} = \bf \delta r,
\label{eq:S_convert}
\eeq
Explicitly, the element of $\mathcal{S}$ is
\beq
(\mathcal{S}_b)_i^\alpha = (\delta_{li} - \delta_{mi}) n_{lm}^\alpha.
\label{eq:S}
\eeq
Here the contact $b$ is formed between particles $m$ and $l$, and $b$ determines the row index.
The column index is determined by $i$ and $\alpha$. Plugging Eq.~(\ref{eq:S}) into Eq.~(\ref{eq:S_convert}), one obtains,
\beq
\left( \delta \vec{R}_i - \delta \vec{R}_j \right) \cdot \vec{n}_{ij} = \delta r_{ij},
\eeq
whose geometric interpretation is transparent: it projects the relative displacement between $i$ and $j$ on to the direction of bond.
On the other hand, $\mathcal{T}$ (a $3N$ by $N_{\rm b}$ matrix) converts  forces $\bf f$ along bonds (a vector of size $N_{\rm b}$)  into the net forces  $\bf F$ on particles (a vector of size $3N$):
\beq
\mathcal{T} {\bf f} = \bf F,
\eeq
or equivalently,
\beq
\sum_{j=1}^{N} c_{ij}  f_{ij} \vec{n}_{ij} = \vec{F}_i,
\eeq
where $c$ is the contact matrix ($c_{ij} = 1$ if particles $i$ and $j$ are in contact, otherwise $c_{ij} = 0$). The matrix $\mathcal{T}$ is the transpose of $\mathcal{S}$,
\beq
\mathcal{T} = \mathcal{S}^{\rm T},
\label{eq:T_S_trans}
\eeq
and their  product is the Hessian matrix,
\beq
\mathcal{H} = \mathcal{T} \mathcal{S}.
\eeq
where $\mathcal{H}$ is defined as,
\beq
\mathcal{H}_{ij}^{\alpha \beta} = \frac{\partial ^2 E}{\partial r_i^\alpha \partial r_j^\beta}.
\eeq

\subsection{Decomposition of the shear modulus into affine and non-affine parts}

A microscopic elasticity theory for amorphous solids has been recently developed~\cite{Maloney2004,  lemaitre2006sum, Karmakar2010a, Zaccone2011, Zaccone2011a}, which decomposes the shear modulus into affine and non-affine parts,
\beq
G = G_{\rm A} - G_{\rm NA}.
\label{eq:G_decomp1}
\eeq
The first term is the seminal Born-Huang formula derived for lattices~\cite{born1954dynamical},
\beq
G_{\rm A} = \frac{1}{N} \sum_{b=1}^{N_{\rm b}} n_{b}^{x} n_b^{z} n_{b}^{x} n_{b}^{z},
\label{eq:GA}
\eeq
considering only affine displacements of particles. Here $N_{\rm b} = NZ/2$
and $\vec{n}_b$ is the unit vector along the bond.

The term $G_{\rm NA}$, which does not appear in lattices, originates from corrections caused by non-affine displacements that are necessary to satisfy the force balance conditions during shear,
\beq
G_{\rm NA} = \frac{1}{N} \sum_{k = 1}^{3 N} \frac{({\bf \Xi}\cdot {\bf v}_k) ({\bf \Xi}\cdot {\bf v}_k)}{\lambda_k}.
\label{eq:GNA}
\eeq
Here $\bf \Xi$ is the {\it affine force} (also called {\it mismatch force}) field acting on $N$ particles; ${\bf v}_k$
and $\lambda_k$ are respectively the $k$-th normalized eigenvector and eigenvalue of the Hessian matrix $\mathcal{H}$.
Both $\bf \Xi$ and ${\bf v}_k$  are vectors of size $3N$, and there are in total $3N$ eigenvectors of the Hessian matrix. The affine force $\bf \Xi$ can be computed explicitly for the given energy Eqs.~(\ref{eq:E}) and (\ref{eq:vij}):
\beq
{\Xi}_{i}^{\alpha xz}  = \frac{\partial ^2 E}{\partial r_i^\alpha \partial \epsilon^{xz}}
 = -\sum_{j} c_{ ij} n_{ ij}^{x}  n_{ ij}^{z} n_{ ij}^\alpha,
\label{eq:Xi}
\eeq
where $\epsilon$ is the strain matrix.

\subsection{Decomposition of the shear modulus into isotropic and anisotropic parts}

A formally different expression of shear modulus is derived by Wyart~\cite{Wyart2005}, based on the duality between force propagation and soft modes. The total shear modulus is decomposed into isotropic and anisotropic parts,
\beq
G = G_{\rm I} + G_{\rm AI},
\eeq
where
\beq
G_{\rm I} = \frac{1}{N}\sum_{b = 1}^{N_{\rm b}} \sum_{p=1}^{\frac{1}{2}N\Delta Z} \tilde{f}_{p, b} \tilde{f}_{p,b} n_{b}^{x}  n_{b}^{z}n_{b}^{x}  n_{b}^{z},
\label{eq:Gz1}
\eeq
and
\beq
G_{\rm AI} =  \frac{1}{N} \sum_{b_1 \neq b_2}^{N_{\rm b}}\sum_{p=1}^{\frac{1}{2}N\Delta Z}   \tilde{f}_{p, b_1} \tilde{f}_{p,b_2} n_{ b_1}^{x}  n_{ b_1}^{z}n_{ b_2}^{x}  n_{ b_2}^{z}.
\label{eq:Gmu1}
\eeq
Here $\tilde{\mathbf{f}}_p$ is {the $p$-th state of self stress,} a normalized vector representing the set of interaction forces (along the bond directions) that can satisfy the  force balance conditions on every particle. The $\frac{1}{2}N\Delta Z$ vectors of $\tilde{\mathbf{f}}_p$ are orthogonal to each other. Among them,  only  $\tilde{\mathbf{f}}_1$ corresponds to the real forces $\mathbf{f}$ generated in simulations, which have to be positive for all contacts:
\beq
\tilde{\mathbf{f}}_1 = \mathbf{f}/\left ( \sum_{b} f_b^2 \right )^{1/2}.
\label{eq:f1}
\eeq
The rest of the vectors $\tilde{\mathbf{f}}_p$ with $p \ge 2$ are "virtual" forces, having a roughly equal number of positive and negative components~\cite{Wyart2005}, {i.e.,  
\beq
\sum_{b=1}^{N_b} \tilde{f}_{p,b} \approx 0
\label{eq:virtual}
\eeq
for $p \ge 2$.}

Under the effective media approximation, i.e., assuming independence between  forces and bond orientations,
$\langle \tilde{f}_{p, b} \tilde{f}_{p,b} n_{b}^{x}  n_{b}^{z}n_{b}^{x}  n_{b}^{z} \rangle \approx \langle \tilde{f}_{p, b} \tilde{f}_{p,b} \rangle\langle n_{b}^{x}  n_{b}^{z}n_{b}^{x}  n_{b}^{z} \rangle$,
Eq.~(\ref{eq:Gz1}) becomes
\beq
G_{\rm I} \approx c_{\rm I} \Delta Z = \frac{1}{30} \Delta Z,
\label{eq:Gz2}
\eeq
where we have used the orthonormality of $\tilde{\mathbf{f}}_p$ and Eq.~(\ref{eq:n4}).
Note that Eq.~(\ref{eq:Gz2})
is independent of bond orientations (or the anisotropy parameter $R_{\rm A}$), thus capturing only isotropic contributions. Indeed, Eq.~(\ref{eq:Gz2}) agrees well with the shear modulus measured in isotropically compressed packings~\cite{Zaccone2011a}, but not in sheared packings (see Fig.~6A).

In Eq.~(\ref{eq:Gmu1}), the term of positive real forces ($p = 1$) dominates; the positive and negative virtual forces are roughly equally probable and therefore the terms of virtual forces ($p \ge 2$) make minor contributions.
The anisotropic shear modulus can then be approximately expressed as,
\beq
G_{\rm AI} \approx \frac{1}{N} \frac{\sum_{ b_1 \neq b_2} f_{b_1} f_{b_2} n_{b_1}^{x}  n_{ b_1}^{z}n_{ b_2}^{x}  n_{ b_2}^{z}}{ \sum_{b} f_b^2},
\label{eq:Gmu2}
\eeq
where we have used Eq.~(\ref{eq:f1}).

\subsection{Equivalence of two decompositions}
Here we  prove the equivalence between the above two decompositions, by deriving Eqs.~(\ref{eq:Gz1}) and~(\ref{eq:Gmu1}) from Eqs.~(\ref{eq:GA}) and (\ref{eq:GNA}). Our derivation is based on
the formalism  developed in Ref.~\cite{Wyart2005}
The goal is to replace the set of eigenvectors ${\bf v}_k$, where $k=1,2,\ldots 3N$, in Eq.~(\ref{eq:GNA}) by the set of equilibrium forces $\tilde{{\bf f}}_p$ that balance all particles, where $p=1,2, \ldots \frac{1}{2}N \Delta Z$. This can be done by imposing the conditions of mechanical equilibrium, which connects displacement fields to equilibrium forces.

The first step is to project the particle-based displacement fields ${\bf v}_k$ onto the directions of bonds, by applying Eq.~(\ref{eq:S_convert}):
\beq
{\bf v}^{\parallel}_k = \frac{1}{\sqrt{\lambda_k}} \mathcal{S} {\bf v}_k,
\label{eq:v_para}
\eeq
where the pre-factor $\frac{1}{\sqrt{\lambda_k}}$ ensures the normalization. Plugging Eq.~(\ref{eq:S}) into the above equation gives,
\beq
{v}^{\parallel}_{k, b} = \frac{1}{\sqrt{\lambda_k}} \sum_{\alpha} \left[\left( v_k \right)_l^\alpha - \left( v_k \right)_m^\alpha \right] n_{lm}^\alpha.
\eeq
Using this expression and Eq.~(\ref{eq:Xi}), we can rewrite Eq.~(\ref{eq:GNA}) as,
\beq
G_{\rm NA} = \frac{1}{N}\sum_{b_2 = 1}^{N_{\rm b}} \sum_{k=1}^{3N} \sum_{b_1 = 1}^{N_{\rm b}} v_{ k,b_1}^{\parallel}v_{ k,b_2}^{\parallel} n_{ b_1}^{x}  n_{ b_1}^{z}n_{ b_2}^{x}  n_{ b_2}^{z}.
\label{eq:GNA2}
\eeq

Second, based on the equilibrium condition $\mathcal{T} \tilde{{\bf f}}_p = 0$,  it can be shown that, for any $k$ and $p$, ${\bf v}^{\parallel}_k$ and $\tilde{{\bf f}}_p$ have to be perpendicular to each other. This is  because,
\beq
{\bf v}^{\parallel}_k \cdot \tilde{{\bf f}}_p =  \frac{1}{\sqrt{\lambda_k}}  \left( {\bf v}_k \right)^{\rm T} \left( \mathcal{T} \tilde{{\bf f}}_p \right) = 0,
\label{eq:v_f_perp}
\eeq
where we have used Eqs.~(\ref{eq:T_S_trans}) and~(\ref{eq:v_para}). Equation~(\ref{eq:v_f_perp}) guarantees the minimization of energy at equilibrium. If we further require  that $\tilde{{\bf f}}_p$'s are perpendicular to each other, then ${\bf v}^{\parallel}_k$ and $\tilde{{\bf f}}_p$ form a complete orthonormal basis of the vector space of dimension $N_{\rm b}$ (note that in the jammed phase, $N_{\rm b} = 3N + \frac{1}{2}N\Delta Z$), which can be represented by an $N_{\rm b}$ by  $N_{\rm b}$ orthonormal matrix,
\beq
\begin{pmatrix}
{v}^{\parallel}_{1, 1} & {v}^{\parallel}_{1, 2} & \cdots & {v}^{\parallel}_{1, N_{\rm b}}\\

{v}^{\parallel}_{2, 1} & {v}^{\parallel}_{2, 2} & \cdots & {v}^{\parallel}_{2, N_{\rm b}} \\

\vdots  \\

{v}^{\parallel}_{3N, 1} & {v}^{\parallel}_{3N, 2} & \cdots & {v}^{\parallel}_{3N, N_{\rm b}} \\

\tilde{f}_{1, 1} & \tilde{f}_{1, 2} & \cdots & \tilde{f}_{1, N_{\rm b}} \\

\tilde{f}_{2, 1} & \tilde{f}_{2, 2} & \cdots & \tilde{f}_{2, N_{\rm b}} \\

\vdots  \\

\tilde{f}_{\frac{1}{2}N\Delta Z, 1} & \tilde{f}_{\frac{1}{2}N\Delta Z, 2} & \cdots & \tilde{f}_{\frac{1}{2}N\Delta Z, N_{\rm b}} \\

\end{pmatrix}
.
\eeq
Because the columns of an  orthonormal  matrix are also  orthonormal, we get,
\beq
\sum_{k=1}^{3N} v_{ k,b_1}^{\parallel} v_{ k,b_2}^{\parallel} + \sum_{p=1}^{\frac{1}{2}N\Delta Z} \tilde{f}_{p, b_1} \tilde{f}_{p,b_2} = \delta_{b_1 b_2}.
\label{eq:v_f_orth}
\eeq

Finally, combining Eq.~(\ref{eq:v_f_orth}) with Eqs.~(\ref{eq:G_decomp1}),~(\ref{eq:GA}) and~(\ref{eq:GNA2}), we obtain Eqs.~(\ref{eq:Gz1}) and~(\ref{eq:Gmu1}).

\subsection{Anisotropic shear modulus}

Next, we relate the anisotropic shear modulus Eq.~(\ref{eq:Gmu2}) to the macroscopic friction coefficient $\mu = \sigma/P$. Using the virial expressions of $\sigma$ and $P$ (see Materials and Methods), we obtain,
\beq
\mu^2 &=& 9\frac{\sum_{b} ( f_b n_b^{x} n_b^{ z})^2 + \sum_{b_1 \neq b_2} f_{b_1} f_{b_2} n_{b_1}^{x} n_{b_1}^{z} n_{b_2}^{x} n_{b_2}^{z}}{\left ( \sum_{b}f_b \right )^2} \nonumber \\
&\approx& 9\frac{\sum_{b_1 \neq b_2} f_{b_1} f_{b_2} n_{b_1}^{x} n_{b_1}^{z} n_{b_2}^{x} n_{b_2}^{z}}{\left ( \sum_{b}f_b \right )^2}.
\label{eq:mu2}
\eeq
Note that the spatial correlation between forces is short-ranged~\cite{OHern2003}, which means that the first term $\sum_{b} ( f_b n_b^{x} n_b^{ z})^2/\left( \sum_{b}f_b \right )^2$ is of order $O(1/N_{\rm b})$, while the second term is of order $O(1)$.
With Eq.~(\ref{eq:mu2}), the anisotropic shear modulus Eq.~(\ref{eq:Gmu2}) can be written as
{\beq
G_{\rm AI} \approx \frac{Z}{18}\frac{\langle f \rangle^2}{\langle f^2 \rangle} \mu^2.
\label{eq:real_mu}
\eeq
}

\begin{figure}
  \centering
  \includegraphics[width=0.9\linewidth]{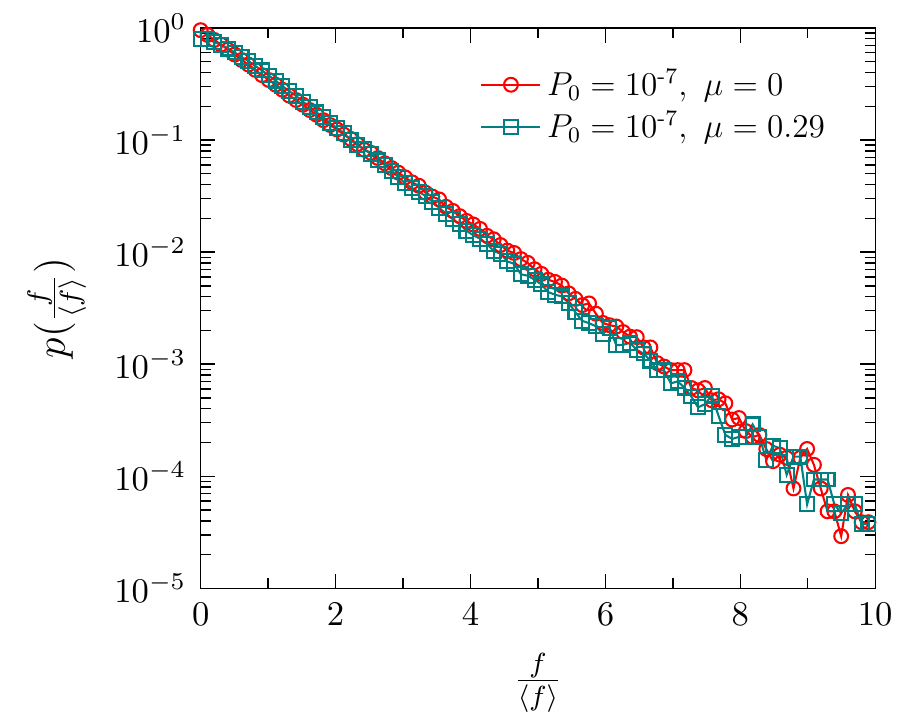}
  \caption{{Distribution of contact force in isotropic and anisotropic systems.} The force distribution functions of isotropic ($P_0 = 10^{-7}, \mu = 0$) and sheared systems ($P_0 = 10^{-7}, \mu = 0.29$).}
  \label{fig:forceHist}
\end{figure}

Our simulation results show that the force distribution $p(f)$ is independent of  $\mu$ (see Fig.~\ref{fig:forceHist}).  From  $p(f)$, it is easy to evaluate $\langle f \rangle^2/\langle f^2 \rangle \approx 0.50$. In addition, $Z \approx 2d = 6$ near the jamming transition. With these values, {when $\Delta Z \to 0$ we have $G_{\rm AI} \approx 0.17 \mu^2$.}


It is also possible to consider the contributions from virtual forces $\tilde{\mathbf{f}}_p$ for  $p \ge 2$.  
In analogy with Eq.~(\ref{eq:mu2}), we obtain, 
\beq
\sum_{b_1 \neq b_2}^{N_{\rm b}}   \tilde{f}_{p, b_1} \tilde{f}_{p,b_2} n_{ b_1}^{x}  n_{ b_1}^{z}n_{ b_2}^{x}  n_{ b_2}^{z} \sim -\mu^2,
\label{eq:virtual_mu}
\eeq
for $p \ge 2$, where we have assumed that $\langle \tilde{f}_{p, b_1} \tilde{f}_{p,b_2} n_{ b_1}^{x}  n_{ b_1}^{z}n_{ b_2}^{x}  n_{ b_2}^{z} 
\rangle \approx \langle \tilde{f}_{p, b_1} \tilde{f}_{p,b_2} \rangle \langle n_{ b_1}^{x}  n_{ b_1}^{z}n_{ b_2}^{x}  n_{ b_2}^{z} 
\rangle$ for any $p$.
The minus sign in Eq.~(\ref{eq:virtual_mu}) comes from the property of virtual forces Eq.~(\ref{eq:virtual}),  according to which,   
$(\sum_b^{N_b} \tilde{f}_{p,b} )^2 = \sum_{b}^{N_b} \tilde{f}_{p_b}^2 + \sum_{b_1 \neq b_2}^{N_b} \tilde{f}_{p,b_1} \tilde{f}_{p,b_2} \approx 0$, and thus $\sum_{b_1 \neq b_2}^{N_b} \tilde{f}_{p,b_1} \tilde{f}_{p,b_2} <0$.
Plugging Eq.~(\ref{eq:virtual_mu}) into  Eq.~(\ref{eq:Gmu1}) gives a negative correction term $\sim -\Delta Z \mu^2$ to the shear modulus Eq.~(\ref{eq:real_mu}). With this correction, $G_{\rm AI}$ becomes,
\beq
G_{\rm AI} = c_{\rm AI}(\Delta Z) \mu^2 = (c_0 - \alpha \Delta Z) \mu^2
\label{eq:g2}
\eeq
where $c_0 = \frac{1}{3}\frac{\langle f  \rangle^2}{\langle f^2 \rangle}$  and $ \alpha$ is a constant that can be determined from the fit of simulation data (see the inset of Fig.~6B). 
Equation~(\ref{eq:g2}) shows that the contribution of virtual forces is higher-order, which only appears in systems above jamming ($\Delta Z  > 0$).

\subsection{Bulk modulus}
It is straightforward to generalize the theoretical analysis to other components of the stiffness tensor, $\mathcal{C}^{\rm \alpha\beta,\gamma\delta}$. Similar to the above derivation, any component $\mathcal{C}^{\rm \alpha\beta,\gamma\delta}$ can be decomposed into isotropic and anisotropic parts. The isotropic part is 
\beq
\mathcal{C}^{\rm \alpha\beta,\gamma\delta}_{\rm I} \approx \frac{1}{2}\langle n_b^{\alpha} n_b^{\beta}n_b^{\gamma} n_b^{\delta} \rangle \Delta Z,
\eeq
which is a generalized formula of Eq.~(\ref{eq:Gz2}),
and the anisotropic part is
\beq
\mathcal{C}^{\rm \alpha\beta,\gamma\delta}_{\rm AI} = \frac{1}{N} \sum_{b_1 \neq b_2}^{N_{\rm b}}\sum_{p=1}^{\frac{1}{2}N\Delta Z}   \tilde{f}_{p, b_1} \tilde{f}_{p,b_2} n_{b_1}^{\alpha}  n_{ b_1}^{\beta}n_{ b_2}^{\gamma}  n_{ b_2}^{\delta},
\eeq
which is a generalized formula of Eq.~(\ref{eq:Gmu1}).
In analogy with the treatment of the shear modulus, we can relate the anisotropic part $\mathcal{C}^{\rm \alpha\beta,\gamma\delta}_{\rm AI}$ to the generalized stress anisotropy, $\mu_{\alpha\beta} = \sigma_{\alpha\beta}/P$, where $\sigma_{\alpha\beta}$ is the $\alpha\beta$-component of stress tensor,
\beq
\mathcal{C}^{\rm \alpha\beta,\gamma\delta}_{\rm AI} = c_{\rm \alpha\beta,\gamma\delta} (\Delta Z) \mu_{\alpha\beta} \mu_{\gamma\delta},
\eeq
where $c_{\rm \alpha\beta,\gamma\delta} (\Delta Z=0) = c_0$ and 
 the sub-leading correction is proportional to $\Delta Z$.
In particular, the bulk modulus $B$ is,
\beq
B = \frac{1}{9}\sum_{\alpha}\sum_{\beta} \mathcal{C}^{\rm \alpha\alpha,\beta\beta} \approx 
c_0 + \left(\frac{1}{18} + \frac{c_0}{6} \right) \Delta Z
\label{eq:bulk_si}
\eeq
which is independent of the  anisotropy.

%

\end{document}